\documentclass[reprint,twocolumn,aps,prd,showpacs]{revtex4-1}
\usepackage{mathbbol}
\usepackage{tensor}
\usepackage[makeroom]{cancel}
\usepackage{times}
\newcommand{\onehalf}{{\textstyle\frac{1}{2}}}
\newcommand{\quarter}{{\textstyle\frac{1}{4}}}
\newcommand{\HC}{\mathcal{H}}
\newcommand{\FC}{\mathcal{F}}
\newcommand{\LC}{\mathcal{L}}

\newcommand{\dd}{\,\mathrm{d}}
\newcommand{\sgn}{\mathrm{sgn}\,}
\newcommand{\pfrac}[2]{\frac{\partial #1}{\partial #2}}
\newcommand{\ppfrac}[3]{\frac{\partial^{2}#1}{\partial #2\partial #3}}
\newcommand{\pppfrac}[4]{\frac{\partial^{3}#1}{\partial #2\partial #3\partial #4}}
\newcommand{\e}{\mathrm{e}}
\begin{document}
\title{General relativity as an extended canonical gauge theory}
\author{J\"urgen Struckmeier}
\email[]{j.struckmeier@gsi.de}
\affiliation{GSI Helmholtzzentrum f\"ur Schwerionenforschung GmbH,
Planckstrasse~1, 64291~Darmstadt, Germany\\
and Goethe University, Max-von-Laue-Strasse~1, 60438~Frankfurt am Main, Germany}
\received{11 February 2015; 
published 22 April 2015 in Phys.~Rev.~D~\textbf{91}, 085030 (2015)}
\begin{abstract}
It is widely accepted that the fundamental geometrical law
of nature should follow from an \emph{action principle}.
The particular subset of transformations of a system's dynamical
variables that maintain the \emph{form\/} of the action principle
comprises the group of \emph{canonical transformations}.
In the context of canonical field theory, the adjective ``extended'' signifies that not
only the \emph{fields\/} but also the \emph{space-time geometry\/} is subject to transformation.
Thus, in order to be physical, the transition to another, possibly
noninertial frame of reference must necessarily constitute an
extended canonical transformation that defines the general mapping
of the \emph{connection coefficients}, hence the quantities that
determine the space-time curvature and torsion of the respective reference frame.
The canonical transformation formalism defines simultaneously the transformation rules
for the \emph{conjugates\/} of the connection coefficients \emph{and\/} for the Hamiltonian.
As will be shown, this yields \emph{unambiguously\/} a particular Hamiltonian
that is form-invariant under the canonical transformation of the connection
coefficients and thus satisfies the general principle of relativity.
This Hamiltonian turns out to be a \emph{quadratic\/} function of the curvature tensor.
Its Legendre-transformed counterpart then establishes a unique
Lagrangian description of the dynamics of space-time that is
\emph{not postulated\/} but \emph{derived from basic principles},
namely the action principle and the general principle of relativity.
Moreover, the resulting theory satisfies the principle of
\emph{scale invariance\/} and is \emph{renormalizable}.
\end{abstract}
\maketitle
\section{\label{sec:intro}Introduction}
The \emph{special\/} principle of relativity states that the fundamental
laws of physics must be form-invariant under Lorentz transformations.
This can be regarded as to require the description of a system to be
form-invariant under a \emph{global\/} transformation group, namely the Lorentz group.
The generalization to noninertial reference frames is
referred to as the \emph{general\/} principle of relativity.
According to this principle, the description of a system is required
to be form-invariant under the corresponding \emph{local\/} transformation group, namely the
\emph{diffeomorphism group\/} that comprises mappings of the local space-time geometry.
In this regard, the transition from the special to the general principle
of relativity meets the \emph{gauge principle}.
The latter requires a physical system that happens to be form-invariant under
a characteristic \emph{global\/} transformation group of the fields to be
rendered form-invariant under the corresponding \emph{local\/} transformation group,
hence the corresponding \emph{explicitly space-time dependent\/} transformation group of the fields.
For this requirement to be met, a set of ``gauge fields'' must be added
to the system's dynamics that obey a specific inhomogeneous transformation rule.
In the case of General Relativity, the local transformation group
is constituted by space-time dependent mappings of the local curvature
and possibly the local torsion of the reference frames.
The ``gauge fields'' are then given by the \emph{connection coefficients\/}---also
referred to as \emph{Christoffel symbols\/} for the particular case of
a coordinate (holonomic) basis of the reference frame.
The connection coefficients also obey a specific inhomogeneous
transformation rule relating different reference frames.

The laws which govern the dynamics of classical systems can be derived from Hamilton's \emph{action principle}.
In this context, a dynamical system is described by a \emph{Lagrangian\/}
or its Legendre transform, the \emph{Hamiltonian}.
From the action principle, the dynamics of a classical particle or of
a classical field can be derived by integrating the Euler-Lagrange
equations or, equivalently, by integrating the \emph{canonical equations}.
The subsequent theory of \emph{canonical transformations\/} then isolates exactly the
subset of those transformations of the dynamical variables that maintain the form of the
action principle---and hence the general form of the \emph{canonical equations}.
As canonical transformations are not restricted to point transformations,
the canonical formalism thus establishes the most general path to work out theories
that are supposed to be form-invariant under the action of a transformation
group of the fields while ensuring the action principle to be maintained.
In the \emph{extended\/} canonical transformation formalism, the space-time
geometry is also subject to transformation.
With the space-time then treated as a dynamic variable, a theory that is form-invariant
as well under the canonical mapping of the connection coefficients then simultaneously
maintains the action principle and satisfies the general principle of relativity.

With this paper, a generalization of the Hamiltonian gauge
transformation formalism is reviewed~\cite{struckmeier13} that extends the
requirement of form-invariance under a \emph{local\/}
transformation group of the fields to also demand form-invariance
under local transformation of the connection coefficients.
Thereby, the well-known formal similarities between non-Abelian
gauge theories~\cite{StrRei12} and general relativity (\cite[p~409]{ryder09},
\cite[p~163]{ChengLi00}) are encountered.
With the transformation rules for the connection coefficients and their
respective canonical conjugates being derived from a \emph{generating function},
it is automatically assured that the extended action principle is
preserved, hence that the actual space-time transformation is \emph{physical}.
No additional assumptions need to be incorporated for setting up
an \emph{amended Hamiltonian\/} that is \emph{locally\/} form-invariant
on the basis of a given \emph{globally}, hence Lorentz-invariant Hamiltonian.
In particular, the connection coefficients are introduced in
the most general way by only specifying their transformation properties.
No \emph{a priori\/} assumptions are incorporated.
In particular, it is \emph{not\/} assumed that the connection coefficients are symmetric
in their lower index pair, hence that a torsion of space-time is excluded~\cite{hehl76}.
Furthermore, following Palatini's approach~\cite{palatini19}, the correlation
of the connection coefficients with the metric emerges from a canonical equation---or
from an Euler-Lagrange equation in the equivalent Lagrangian description---rather than being postulated.

Prior to working out the general local space-time transformation theory
in the extended canonical formalism in Sec.~\ref{sec:example-space-time-trans-curv},
the formalism of extended Lagrangians and Hamiltonians and their subsequent field
equations is presented in Secs.~\ref{sec:gen-lagr} and~\ref{sec:gen-ham}.
With the space-time treated as a dynamical variable, the extended
Lagrangians and Hamiltonians are defined to also depend on the
connection coefficients and their respective conjugates.
The general space-time transformation theory in classical vacuum
is then based on an extended generating function that defines the
mapping of the connection coefficients in the transition
from one frame of reference to another.
As an extended generating function \emph{simultaneously\/} defines
the transformation rule for the canonical conjugates of the connection
coefficients as well as the transformation law for the extended Hamiltonians,
one directly encounters a particular extended Hamiltonian that
is \emph{form-invariant\/} under the required transformation law of the
connection coefficients while maintaining the action principle.
The set of canonical field equations following from the obtained
gauge-invariant Hamiltonian now establishes a field equation
for the Riemann curvature tensor that is no longer \emph{postulated},
but uniquely emerges from both the action principle and the general principle of relativity.
\section{\label{sec:gen-lagr}Extended Lagrangians $\LC_{\e}$
in the realm of classical field theory}
\subsection{Variational principle, extended set
of Euler-Lagrange field equations}
Similar to point dynamics, the Lagrangian formulation of continuum dynamics
(see, e.g., \cite{saletan98}) is based on a scalar Lagrange function $\LC$ that
is supposed to contain the complete information on the given physical system.
In a first-order scalar field theory, the Lagrangian $\LC$
is defined to depend on $I=1,\ldots,N$---possibly interacting---scalar
fields $\phi_{I}(x)$, on the vector of independent space-time variables
$x^{\mu}$, and on the first derivatives of the scalar fields $\phi_{I}$
with respect to the independent variables, i.e., on the covariant vectors ($1$-forms)
\begin{displaymath}
\pfrac{\phi_{I}}{x^{\nu}}\equiv
\left(\pfrac{\phi_{I}}{x^{0}},
\pfrac{\phi_{I}}{x^{1}},\pfrac{\phi_{I}}{x^{2}},\pfrac{\phi_{I}}{x^{3}}\right).
\end{displaymath}
The Euler-Lagrange field equations are then obtained
as the zero of the variation $\delta S$ of the action functional over a space-time region $R$
\begin{equation}\label{action-int}
S=\int_{R}\LC\left(\phi_{I},\pfrac{\phi_{I}}{x^{\nu}},x^{\mu}\right)\dd^{4}x,\qquad
\delta S\stackrel{!}{=}0
\end{equation}
as
\begin{equation}\label{elgl}
\pfrac{}{x^{j}}\pfrac{\LC}{\left(\pfrac{\phi_{I}}
{x^{j}}\right)}-\pfrac{\LC}{\phi_{I}}=0.
\end{equation}
If the Lagrangian $\LC$ is to describe the dynamics of a set of
\emph{vector fields\/} $a_{I}^{\mu}(x)$ that possibly couple to the scalar fields $\phi_{I}$,
then the additional Euler-Lagrange equations take on the similar form
\begin{equation}\label{elgl-ext1-vec}
\pfrac{}{x^{j}}\pfrac{\LC}{\left(\pfrac{a_{I}^{\mu}}
{x^{j}}\right)}-\pfrac{\LC}{a_{I}^{\mu}}=0,\quad
\LC=\LC\!\left(\phi_{I},\pfrac{\phi_{I}}{x^{\nu}},a_{I}^{\mu},\pfrac{a_{I}^{\mu}}{x^{\nu}},x^{\mu}\!\right)\!\!.
\end{equation}
The derivatives of vectors do \emph{not\/} transform as tensors.
This means that the Euler-Lagrange equations~(\ref{elgl}) and~(\ref{elgl-ext1-vec})
are not form-invariant under transformations of the space-time metric.
Any field equation emerging from Eqs.~(\ref{elgl})
and~(\ref{elgl-ext1-vec}) that holds a local frame $y$ must finally
be rendered a \emph{tensor equation\/} in order for the theory
described by $\LC$ to hold in any reference frame.
This is achieved by converting all \emph{partial derivatives\/}
in the field equations into \emph{covariant derivatives}.

In analogy to the extended formalism of point mechanics
(\cite{struckmeier05,struckmeier09}),
the action integral from Eq.~(\ref{action-int}) can directly be cast
into a more general form by \emph{decoupling\/} its integration measure
from a possibly explicit $x^{\mu}$-dependence of the Lagrangian $\LC$
\begin{equation}\label{action-int1}
S=\int_{R^{\prime}}\LC\left(\phi_{I},\pfrac{\phi_{I}}{x^{\nu}},x^{\mu}\right)\det\Lambda\dd^{4}y.
\end{equation}
Herein, $\det\Lambda\neq0$ stands for the determinant of the Jacobi
matrix $\Lambda=(\Lambda_{\nu^{\prime}}^{\mu})$ that is associated with a
regular transformation $x^{\mu}\mapsto y^{\mu}$ of the independent variables
and the corresponding transformation $R\mapsto R^{\prime}$ of the integration region
\begin{eqnarray}
\Lambda&=&(\Lambda_{\nu^{\prime}}^{\mu})=\left(\begin{array}{ccc}
\pfrac{x^{0}}{y^{0}}&\ldots&\pfrac{x^{0}}{y^{3}}\\
\vdots&\ddots&\vdots\\
\pfrac{x^{3}}{y^{0}}&\ldots&\pfrac{x^{3}}{y^{3}}\end{array}\right),\nonumber\\
\det\Lambda&=&\pfrac{(x^{0},\ldots,x^{3})}{(y^{0},\ldots,y^{3})}\neq0\nonumber\\
\Lambda_{\nu^{\prime}}^{\mu}(y)&=&\pfrac{x^{\mu}(y)}{y^{\nu}},\qquad
\Lambda_{\nu}^{\mu^{\prime}}(x)=\pfrac{y^{\mu}(x)}{x^{\nu}},\nonumber\\
\Lambda_{\nu^{\prime}}^{\alpha}\Lambda_{\alpha}^{\mu^{\prime}}&=&
\Lambda_{\alpha^{\prime}}^{\mu}\Lambda_{\nu}^{\alpha^{\prime}}=\delta_{\nu}^{\mu}.
\label{jacdet}
\end{eqnarray}
Here, the prime indicates the location of the new independent variables, $y^{\mu}$.
As this transformation constitutes a mapping of the space-time
metric, the $\Lambda_{\nu^{\prime}}^{\mu}$ are referred to
as the \emph{space-time coefficients}.
The integrand of Eq.~(\ref{action-int1}) can be thought of as defining
an \emph{extended\/} Lagrangian $\LC_{\e}$,
\begin{eqnarray}
&&\LC_{\e}\left(\phi_{I},\pfrac{\phi_{I}(y)}{y^{\nu}},x^{\mu}(y),
\pfrac{x^{\mu}(y)}{y^{\nu}}\right)\nonumber\\
&&\quad=\LC\left(\phi_{I},\pfrac{\phi_{I}(y)}{y^{k}}
\pfrac{y^{k}}{x^{\nu}},x^{\mu}(y)\right)\det\Lambda.
\label{L1-def}
\end{eqnarray}
In the language of tensor calculus, the conventional Lagrangian $\LC$ represents
an \emph{absolute scalar\/} whereas the extended Lagrangian $\LC_{\e}$ transforms
as a \emph{relative scalar of weight $w=1$\/} under a mapping of the independent variables, $y^{\mu}$.
With this property, $\LC_{\e}$ is referred to as a \emph{scalar density}.
Both, $\LC$ and $\LC_{\e}$ have the dimension of $\mathrm{Length}^{-4}$ in natural units ($c=1$).
The now \emph{dependent\/} variables $x^{0},\ldots,x^{3}$ in
the argument list of $\LC_{\e}$ can be regarded as an \emph{extension\/}
of the set of fields $\phi_{I}, I=1,\ldots,N$.
In other words, the $x^{\mu}(y)$ are treated on equal footing
with the fields $\phi_{I}(y)$.
In terms of the extended Lagrangian $\LC_{\e}$, the action integral
over $\dd^{4}y$ from Eq.~(\ref{action-int1}) is converted into an
integral over an \emph{autonomous\/} Lagrangian, hence over a Lagrangian
that does not \emph{explicitly\/} depend on its independent variables, $y^{\mu}$
\begin{equation}\label{action-int2}
S=\int_{R^{\prime}}\LC_{\e}\left(\phi_{I},
\pfrac{\phi_{I}}{y^{\nu}},x^{\mu}(y),\pfrac{x^{\mu}}{y^{\nu}}\right)\dd^{4}y.
\end{equation}
As this action integral has exactly the form of the initial
one from Eq.~(\ref{action-int}), the Euler-Lagrange field equations
emerging from the variation of Eq.~(\ref{action-int2}) take on form
of Eq.~(\ref{elgl}) (see Appendix~\ref{app3})
\begin{equation}\label{elgl-ext1}
\pfrac{}{y^{j}}\pfrac{\LC_{\e}}{\left(\pfrac{\phi_{I}}
{y^{j}}\right)}-\pfrac{\LC_{\e}}{\phi_{I}}=0,\qquad
\pfrac{}{y^{j}}\pfrac{\LC_{\e}}{\left(\pfrac{x^{\mu}}
{y^{j}}\right)}-\pfrac{\LC_{\e}}{x^{\mu}}=0.
\end{equation}
An extended Lagrangian $\LC_{\e}$ that is correlated to a conventional
Lagrangian $\LC$ according to Eq.~(\ref{L1-def}) is referred to as a
\emph{trivial\/} extended Lagrangian.
Clearly, multiplying a conventional Lagrangian $\LC$ by $\det\Lambda$
and expressing the $x^{\nu}$-derivatives of the fields by means of the
chain rule in terms of $y^{\nu}$-derivatives does not add any information.
The system's description in terms of a trivial $\LC_{\e}$ is thus
\emph{equivalent\/} to that provided by $\LC$.

Yet, a dynamical system that exhibits a dynamical space-time is generally described by an
extended Lagrangian $\LC_{\e}$ that does \emph{not\/} have a conventional counterpart $\LC$.
Furthermore, is possible to define extended Lagrangians $\LC_{\e}$ that depend in addition
on the \emph{connection coefficients\/} $\gamma\indices{^{\eta}_{\alpha\xi}}(x)$
(see Appendix~\ref{app2}) and their respective $x^{\nu}$-derivatives that describe the
space-time curvature in the $x$ reference frame
\begin{eqnarray}
S&=&\int_{R^{\prime}}\LC_{\e}\left(\phi_{I},\pfrac{\phi_{I}}{y^{\nu}},
\gamma\indices{^{\eta}_{\alpha\xi}}(x),\pfrac{\gamma\indices{^{\eta}_{\alpha\xi}}}{x^{\nu}},
x^{\mu}(y),\pfrac{x^{\mu}}{y^{\nu}}\right)\!\!\dd^{4}y,\nonumber\\
\delta S&\stackrel{!}{=}&0.
\label{action-int3}
\end{eqnarray}
This description follows the path of A.~Palatini~\cite{palatini19,janssen08}, who treated
the connection coefficients and the metric as \emph{a priori\/} independent quantities.
Their correlation then follows from the extended Lagrangian in Eq.~(\ref{action-int3})
by means of the additional Euler-Lagrange equation (see Appendix~\ref{app3})
\begin{equation}\label{elgl-ext2}
\pfrac{}{x^{j}}\pfrac{\LC_{\e}}{\left(\pfrac{\gamma\indices{^{\eta}_{\alpha\xi}}}
{x^{j}}\right)}-\pfrac{\LC_{\e}}{\gamma\indices{^{\eta}_{\alpha\xi}}}=0.
\end{equation}
Equations~(\ref{elgl-ext1}) and (\ref{elgl-ext2}) must be simultaneously fulfilled
in order to minimize the action functional~(\ref{action-int3}).
This determines uniquely the system's dynamics which includes the
dynamics of the space-time geometry.
\section{\label{sec:gen-ham}Extended
Hamiltonians $\HC_{\e}$ in classical field theory}
\subsection{Extended canonical field equations}
For a covariant Hamiltonian description, the momentum fields $\pi_{I}^{\nu}$ and
$\tilde{\pi}_{I}^{\nu}$ must be defined as the \emph{dual quantities\/} of the
derivatives of the fields $\phi_{I}$ according to
\begin{equation}\label{p1-def}
\pi_{I}^{\nu}(x)=\pfrac{\LC}{\left(\pfrac{\phi_{I}}{x^{\nu}\vphantom{y^{\nu}}}\right)},\qquad
\tilde{\pi}_{I}^{\nu}(y)=\pfrac{\LC_{\e}}{\left(\pfrac{\phi_{I}}{y^{\nu}}\right)}.
\end{equation}
The momentum fields $\tilde{\pi}_{I}^{\nu}$ emerging from the extended
Lagrangian density $\LC_{\e}$ transform as
\begin{equation}\label{p1-def1}
\tilde{\pi}_{I}^{\nu}(y)=\pi_{I}^{j}(x)\,\pfrac{y^{\nu}}{x^{j}}\,\det\Lambda
\,\Leftrightarrow\,
\pi_{I}^{\nu}(x)=\tilde{\pi}_{I}^{j}(y)\,\pfrac{x^{\nu}}{y^{j}}\,
\frac{1}{\det\Lambda}.
\end{equation}
As indicated by the tilde, the $\tilde{\pi}_{I}^{\nu}=\pi_{I}^{\nu}\det\Lambda$
represent \emph{tensor densities of weight $w=1$}, whereas the $\pi_{I}^{\nu}$
transform as \emph{absolute\/} tensors.

Corresponding to the momentum field $\tilde{\pi}_{I}^{\nu}(y)$ that
constitutes the dual counterpart of the Lagrangian dynamical variable
$\partial\phi_{I}(y)/\partial y^{\nu}$, the canonical variable
$\tilde{t}\indices{_{\mu}^{\nu}}$ is defined formally as the
dual counterpart of $\partial x^{\mu}/\partial y^{\nu}$ and
$\tilde{r}\indices{_{\eta}^{\alpha\xi\nu}}(x)$ as the dual
counterpart of $\partial\gamma\indices{^{\eta}_{\alpha\xi}}(x)/\partial x^{\nu}$.
Thus, $\tilde{t}\indices{_{\mu}^{\nu}}$ and $\tilde{r}\indices{_{\eta}^{\alpha\xi\nu}}(x)$
follow similarly to Eq.~(\ref{p1-def}) from the respective partial derivative of the extended
Lagrangian density $\LC_{\e}$ as
\begin{eqnarray}
\tilde{t}\indices{_{\mu}^{\nu}}&=&-\pfrac{\LC_{\e}}{\left(
\pfrac{x^{\mu}(y)}{y^{\nu}}\right)}
=\tilde{t}\indices{_{j}^{\nu}}(y)\,\pfrac{y^{j}}{x^{\mu}}=
\tilde{t}\indices{_{\mu}^{j}}(x)\,\pfrac{y^{\nu}}{x^{j}},\nonumber\\
\tilde{r}\indices{_{\eta}^{\alpha\xi\nu}}(x)&=&\pfrac{\LC_{\e}}{\left(
\pfrac{\gamma\indices{^{\eta}_{\alpha\xi}}(x)}{x^{\nu}}\right)}.
\label{pext-def}
\end{eqnarray}
The nontensorial quantity $\gamma\indices{^{\eta}_{\alpha\xi}}(x)$
must be derived with respect to the $x^{\nu}$ rather than with respect to $y^{\nu}$.
Its canonical conjugate $\tilde{r}\indices{_{\eta}^{\alpha\xi\nu}}(x)$ then
also refers to the space-time event $x$.
Possible symmetry properties in $\eta,\alpha,\xi$ of
$\partial\gamma\indices{^{\eta}_{\alpha\xi}}/\partial x^{\nu}$
must agree with those of $\tilde{r}\indices{_{\eta}^{\alpha\xi\nu}}$
in order for both quantities to be actually dual to each other.
The Hamiltonian $\HC$ and the correlated extended Hamiltonian $\HC_{\e}$ can now be defined as the
\emph{covariant Legendre transform\/} of the Lagrangian $\LC$ and the pertaining extended Lagrangian $\LC_{\e}$
\begin{eqnarray}
&&\HC\big(\phi_{I},\pi_{I},\gamma,r,x\big)\nonumber\\
&&\quad=\pi_{J}^{j}\pfrac{\phi_{J}}{x^{j}}+r\indices{_{\eta}^{\alpha\xi j}}
\pfrac{\gamma\indices{^{\eta}_{\alpha\xi}}}{x^{j}}\nonumber\\
&&\qquad\mbox{}-\LC\left(\phi_{I},\pfrac{\phi_{I}}{x^{\nu}},\gamma\indices{^{\eta}_{\alpha\xi}},
\pfrac{\gamma\indices{^{\eta}_{\alpha\xi}}}{x^{\nu}},x^{\mu}\right)\nonumber\\
&&\HC_{\e}\big(\phi_{I},\tilde{\pi}_{I},\gamma,\tilde{r},x,\tilde{t}\,\big)\nonumber\\
&&\quad=\tilde{\pi}_{J}^{j}\pfrac{\phi_{J}}{y^{j}}+\tilde{r}\indices{_{\eta}^{\alpha\xi j}}
\pfrac{\gamma\indices{^{\eta}_{\alpha\xi}}}{x^{j}}-
\tilde{t}\indices{_{\alpha}^{\beta}}\pfrac{x^{\alpha}}{y^{\beta}}\nonumber\\
&&\qquad\mbox{}-\LC_{\e}\left(\phi_{I},\pfrac{\phi_{I}}{y^{\nu}},
\gamma\indices{^{\eta}_{\alpha\xi}},\pfrac{\gamma\indices{^{\eta}_{\alpha\xi}}}{x^{\nu}},
x^{\mu},\pfrac{x^{\mu}}{y^{\nu}}\right).
\label{H1-def1}
\end{eqnarray}
The correlation~(\ref{L1-def}) of conventional and extended Lagrangians thus entails
a corresponding correlation of conventional and extended Hamiltonians,
\begin{equation}\label{H-H1-coor}
\HC_{\e}=\HC\det\Lambda-\tilde{t}\indices{_{\alpha}^{\beta}}\pfrac{x^{\alpha}}{y^{\beta}}.
\end{equation}
With this definition of the extended Hamiltonian $\HC_{\e}$,
one encounters an extended set of the \emph{canonical equations}.
Calculating explicitly the partial derivatives of $\HC_{\e}$
from Eq.~(\ref{H1-def1}) with respect to all canonical variables yields
\begin{eqnarray}
\pfrac{\HC_{\e}}{\tilde{\pi}_{I}^{\nu}}&=&
\pfrac{\tilde{\pi}_{J}^{j}}{\tilde{\pi}_{I}^{\nu}}\pfrac{\phi_{J}}{y^{j}}=\delta_{IJ}
\delta_{\nu}^{j}\pfrac{\phi_{J}}{y^{j}}=\pfrac{\phi_{I}}{y^{\nu}}\nonumber\\
\pfrac{\HC_{\e}}{\tilde{r}\indices{_{\eta}^{\alpha\xi\nu}}}&=&
\pfrac{\tilde{r}\indices{_{\eta}^{\alpha\xi j}}}{\tilde{r}\indices{_{\eta}^{\alpha\xi\nu}}}
\pfrac{\gamma\indices{^{\eta}_{\alpha\xi}}}{x^{j}}=
\delta_{\nu}^{j}\pfrac{\gamma\indices{^{\eta}_{\alpha\xi}}}{x^{j}}=
\pfrac{\gamma\indices{^{\eta}_{\alpha\xi}}}{x^{\nu}}\nonumber\\
\pfrac{\HC_{\e}}{\tilde{t}\indices{_{\mu}^{\nu}}}&=&
-\pfrac{\tilde{t}\indices{_{i}^{j}}}{\tilde{t}\indices{_{\mu}^{\nu}}}\pfrac{x^{i}}{y^{j}}=
-\delta_{\nu}^{j}\delta_{i}^{\mu}\pfrac{x^{i}}{y^{j}}=
-\pfrac{x^{\mu}}{y^{\nu}}\nonumber\\
\pfrac{\HC_{\e}}{\phi_{I}}&=&-\pfrac{\LC_{\e}}{\phi_{I}}=
-\pfrac{}{y^{j}}\pfrac{\LC_{\e}}{\left(\pfrac{\phi_{I}}{y^{j}}\right)}=
-\pfrac{\tilde{\pi}_{I}^{j}}{y^{j}}\nonumber\\
\pfrac{\HC_{\e}}{\gamma\indices{^{\eta}_{\alpha\xi}}}&=&
-\pfrac{\LC_{\e}}{\gamma\indices{^{\eta}_{\alpha\xi}}}=
-\pfrac{}{x^{j}}\pfrac{\LC_{\e}}{\left(\pfrac{\gamma\indices{^{\eta}_{\alpha\xi}}}
{x^{j}}\right)}=-\pfrac{\tilde{r}\indices{_{\eta}^{\alpha\xi j}}}{x^{j}}\nonumber\\
\pfrac{\HC_{\e}}{x^{\mu}}&=&-\pfrac{\LC_{\e}}{x^{\mu}}=
-\pfrac{}{y^{j}}\pfrac{\LC_{\e}}{\left(\pfrac{x^{\mu}}
{y^{j}}\right)}=\pfrac{\tilde{t}\indices{_{\mu}^{j}}}{y^{j}}.
\label{ext-fgln}
\end{eqnarray}
Obviously, an extended Hamiltonian $\HC_{\e}$---through its $\phi_{I}$,
$\gamma\indices{^{\eta}_{\alpha\xi}}$, and $x^{\mu}$ dependencies---only
determines the \emph{divergences\/} $\partial\tilde{\pi}_{I}^{j}/\partial y^{j}$,
$\partial\tilde{r}\indices{_{\eta}^{\alpha\xi j}}/\partial x^{j}$,
and $\partial\tilde{t}\indices{_{\mu}^{j}}/\partial y^{j}$
but \emph{not\/} the individual components $\tilde{\pi}_{I}^{\nu}$,
$\tilde{r}\indices{_{\eta}^{\alpha\xi\nu}}$, and $\tilde{t}\indices{_{\mu}^{\nu}}$
of the canonical ``momentum'' tensor densities.
Consequently, the momenta are only determined by the Hamiltonian $\HC_{\e}$
up to divergence-free functions.

The action integral from Eq.~(\ref{action-int3}) can be equivalently
expressed in terms of the extended Hamiltonian $\HC_{\e}$
by applying the Legendre transform~(\ref{H1-def1})
\begin{eqnarray}
S&=&\int_{R^{\prime}}\bigg(
\tilde{\pi}_{J}^{j}\pfrac{\phi_{J}}{y^{j}}+
\tilde{r}\indices{_{\eta}^{\alpha\xi j}}
\pfrac{\gamma\indices{^{\eta}_{\alpha\xi}}}{x^{j}}-
\tilde{t}\indices{_{i}^{j}}\pfrac{x^{i}}{y^{j}}\nonumber\\
&&\qquad\;\mbox{}-\HC_{\e}\left(\phi_{I},\tilde{\pi}_{I},\gamma,
\tilde{r},x,\tilde{t}\,\right)\bigg)\dd^{4}y.
\label{action-int4}
\end{eqnarray}
This representation of the action integral forms the basis on which the
extended canonical transformation formalism will be worked out in Sec.~\ref{sec:gen-ct}.

In case that the extended Lagrangian describes the dynamics of (covariant) vector fields,
$a_{I\mu}(y)$, the canonical momentum fields are to be defined as
\begin{equation}\label{p1-def-vec}
\tilde{p}_{I}^{\,\mu\nu}(y)=\pfrac{\LC_{\e}}{\left(\pfrac{a_{I\mu}}{y^{\nu}}\right)}.
\end{equation}
Similar to Eq.~(\ref{p1-def1}), the tensor densities $\tilde{p}_{I}^{\,\mu\nu}$
then represent the \emph{dual quantities\/} of the $y^{\nu}$-derivatives of the vector
fields $a_{I\mu}$ and hence the \emph{canonical conjugates\/} of the $a_{I\mu}(y)$
in the extended covariant Hamiltonian description.
\section{\label{sec:gen-ct}Extended canonical transformations}
\subsection{Generating function of type $\FC_{1}$}
In the realm of classical field theory, the condition for defining extended canonical
transformations that include mappings $x^{\mu}(y)\mapsto X^{\mu}(y)$ of the respective
space-time reference systems $x^{\mu}$ and $X^{\mu}$ and of the connection coefficients
$\gamma\indices{^{\eta}_{\alpha\xi}}(x)\mapsto\Gamma\indices{^{\eta}_{\alpha\xi}}(X)$
is based on the extended version of the action functional~(\ref{action-int4}).
Specifically, the action principle $\delta S\stackrel{!}{=}0$ is again
required to be conserved under the action of the transformation, with
$y^{\mu}$ denoting the common independent variables of both reference frames.
Requiring the variation of the action functional to be maintained implies the
\emph{integrand\/} to be determined only up to the divergence of a $4$-vector function
$\FC_{1}^{\mu}=\FC_{1}^{\mu}(\phi_{I},\Phi_{I},\gamma,\Gamma,x,X)$ on the original and the
transformed dynamical ``field'' variables.
The \emph{generating function\/} $\FC_{1}^{\mu}$ depends on the sets of
original fields $\phi_{I}$ and transformed fields $\Phi_{I}$ in conjunction
with the connection coefficients $\gamma\indices{^{\eta}_{\alpha\xi}}(x)$,
$\Gamma\indices{^{\eta}_{\alpha\xi}}(X)$ at the original space-time events
$x^{\nu}(y)$ and transformed ones, $X^{\nu}(y)$
\begin{eqnarray}
&&\quad\LC_{\e}=\LC_{\e}^{\prime}+\pfrac{\FC_{1}^{j}}{y^{j}}
\qquad\Longleftrightarrow\nonumber\\
&&\tilde{\pi}_{J}^{j}\pfrac{\phi_{J}}{y^{j}}+
\tilde{r}\indices{_{\eta}^{\alpha\xi k}}\pfrac{\gamma\indices{^{\eta}_{\alpha\xi}}}{x^{k}}-
\tilde{t}\indices{_{\alpha}^{\beta}}\pfrac{x^{\alpha}}{y^{\beta}}-\HC_{\e}\nonumber\\
&&\quad=\tilde{\Pi}_{J}^{j}\pfrac{\Phi_{J}}{y^{j}}+
\tilde{R}\indices{_{\eta}^{\alpha\xi k}}\pfrac{\Gamma\indices{^{\eta}_{\alpha\xi}}}{X^{k}}-
\tilde{T}\indices{_{\alpha}^{\beta}}\pfrac{X^{\alpha}}{y^{\beta}}-
\HC_{\e}^{\prime}+\pfrac{\FC_{1}^{j}}{y^{j}}.\nonumber\\
\label{eq:intbed-ext}
\end{eqnarray}
At this point, the tensor densities $\tilde{r}\indices{_{\eta}^{\alpha\xi\mu}}(x)$ and
$\tilde{R}\indices{_{\eta}^{\alpha\xi\mu}}(X)$ merely denote formally the respective
canonical conjugates of the connection coefficients, $\gamma\indices{^{\eta}_{\alpha\xi}}(x)$
and $\Gamma\indices{^{\eta}_{\alpha\xi}}(X)$, whose dynamics---and hence whose
physical meaning---follows later from the corresponding canonical equation.
Since the independent variables $y^{\mu}$ are \emph{not\/} transformed,
the divergence of the vector function
$\FC_{1}^{\mu}(\phi_{I},\Phi_{I},\gamma,\Gamma,x,X)$
can be expressed \emph{locally\/} as the ordinary divergence
\begin{eqnarray}
\pfrac{\FC_{1}^{j}}{y^{j}}&=&
\pfrac{\FC_{1}^{j}}{\phi_{I}}\pfrac{\phi_{I}}{y^{j}}+
\pfrac{\FC_{1}^{j}}{\Phi_{I}}\pfrac{\Phi_{I}}{y^{j}}+
\pfrac{\FC_{1}^{j}}{\gamma\indices{^{\eta}_{\alpha\xi}}}\pfrac{x^{k}}{y^{j}}
\pfrac{\gamma\indices{^{\eta}_{\alpha\xi}}}{x^{k}}\nonumber\\
&&\mbox{}+\pfrac{\FC_{1}^{j}}{\Gamma\indices{^{\eta}_{\alpha\xi}}}\pfrac{X^{k}}{y^{j}}
\pfrac{\Gamma\indices{^{\eta}_{\alpha\xi}}}{X^{k}}+\pfrac{\FC_{1}^{j}}{x^{i}}\pfrac{x^{i}}{y^{j}}+
\pfrac{\FC_{1}^{j}}{X^{i}}\pfrac{X^{i}}{y^{j}}.\nonumber\\
\label{eq:divF-ext}
\end{eqnarray}
Comparing the coefficients of~Eqs.~(\ref{eq:intbed-ext}) and~(\ref{eq:divF-ext}),
the \emph{extended\/} local coordinate representation of the transformation
rules induced by the extended generating function $\FC_{1}^{\mu}(y)$ are
\begin{eqnarray}
\tilde{\pi}_{I}^{\mu}&=&\pfrac{\FC_{1}^{\mu}}{\phi_{I}},\quad
\tilde{\Pi}_{I}^{\mu}=-\pfrac{\FC_{1}^{\mu}}{\Phi_{I}},\quad
\tilde{r}\indices{_{\eta}^{\alpha\xi\mu}}=
\pfrac{\FC_{1}^{j}}{\gamma\indices{^{\eta}_{\alpha\xi}}}\pfrac{x^{\mu}}{y^{j}},\nonumber\\
\tilde{R}\indices{_{\eta}^{\alpha\xi\mu}}&=&
-\pfrac{\FC_{1}^{j}}{\Gamma\indices{^{\eta}_{\alpha\xi}}}\pfrac{X^{\mu}}{y^{j}},\quad
\tilde{t}\indices{_{\nu}^{\mu}}=-\pfrac{\FC_{1}^{\mu}}{x^{\nu}},\quad
\tilde{T}\indices{_{\nu}^{\mu}}=\pfrac{\FC_{1}^{\mu}}{X^{\nu}},\nonumber\\
\HC_{\e}^{\prime}&=&\HC_{\e},\;
\HC^{\prime}\det\Lambda^{\prime}-\tilde{T}\indices{_{\alpha}^{\beta}}\pfrac{X^{\alpha}}{y^{\beta}}=
\HC\det\Lambda-\tilde{t}\indices{_{\alpha}^{\beta}}\pfrac{x^{\alpha}}{y^{\beta}}.\nonumber\\
\label{eq:genF1-ext}
\end{eqnarray}
Note that the nontensor quantities $\gamma\indices{^{\eta}_{\alpha\xi}}(x)$
must always refer to the original reference system, $x$, and, correspondingly,
$\Gamma\indices{^{\eta}_{\alpha\xi}}(X)$ to the transformed reference system, $X$.
The reference systems of their local canonical conjugates, hence the tensor densities
$\tilde{r}\indices{_{\eta}^{\alpha\xi\mu}}(x)$ and
$\tilde{R}\indices{_{\eta}^{\alpha\xi\mu}}(X)$ are defined accordingly.
The indices of $\tilde{t}\indices{_{\nu}^{\mu}}$ and
$\tilde{T}\indices{_{\nu}^{\mu}}$ in Eq.~(\ref{eq:genF1-ext})
are related to different reference systems.
The upper index of $\tilde{t}\indices{_{\nu}^{\mu}}$ and
$\tilde{T}\indices{_{\nu}^{\mu}}$ pertains to the reference system $y$,
whereas the lower index refers to the reference systems $x$ and $X$, respectively.
In order to attribute these quantities a definite space-time location, hence
to convert them into regular tensors, the transformations follow as
\begin{eqnarray*}
\tilde{\theta}\indices{_{\nu}^{\mu}}(x)&=&
\tilde{t}\indices{_{\nu}^{j}}\pfrac{x^{\mu}}{y^{j}}=
-\pfrac{\FC_{1}^{j}}{x^{\nu}}\pfrac{x^{\mu}}{y^{j}},\nonumber\\
\tilde{\Theta}\indices{_{\nu}^{\mu}}(X)&=&
\tilde{T}\indices{_{\nu}^{j}}\pfrac{X^{\mu}}{y^{j}}=
\pfrac{\FC_{1}^{j}}{X^{\nu}}\pfrac{X^{\mu}}{y^{j}},
\end{eqnarray*}
in agreement with the definition of $\tilde{t}\indices{_{\nu}^{\mu}}$ as the
\emph{dual quantity\/} of $\partial x^{\nu}/\partial y^{\mu}$ from Eq.~(\ref{pext-def}).
As shown in Appendix~\ref{app1}, the tensor $\theta\indices{_{\nu}^{\mu}}(x)$
represents the canonical energy-momentum tensor of the original system
if the dynamical system is described by a \emph{trivial\/} extended
Lagrangian $\LC_{\e}$ or its corresponding Legendre-transform that
defines a trivial extended Hamiltonian, $\HC_{\e}$.

According to Eqs.~(\ref{eq:genF1-ext}), the \emph{value\/} of an extended
Hamiltonian $\HC_{\e}$ is conserved under extended canonical transformations.
Hence, the transformed extended Hamiltonian density
$\HC_{\e}^{\prime}(\Phi_{I},\tilde{\Pi}_{I},\Gamma,\tilde{R},X,\tilde{T}\,)$
is obtained by expressing the original extended Hamiltonian
$\HC_{\e}(\phi_{I},\tilde{\pi}_{I},\gamma,\tilde{r},x,\tilde{t}\,)$
in terms of the transformed fields $\Phi_{I}$, $\tilde{\Pi}_{I}^{\mu}$,
the transformed $\Gamma\indices{^{\eta}_{\alpha\xi}}$, $\tilde{R}\indices{_{\eta}^{\alpha\xi\mu}}$
and the transformed space-time location $X^{\nu}$, and its canonical conjugate,
$\tilde{T}\indices{_{\nu}^{\mu}}$.
\subsection{Generating function of type $\FC_{2}$}
The generating function of an extended canonical transformation
can alternatively be expressed in terms of a function $\FC_{2}^{\mu}$ of
the original fields $\phi_{I}$, the connection coefficients of the original
system, $\gamma\indices{^{\eta}_{\alpha\xi}}(x)$, and the original space-time
coordinates, $x^{\mu}$ in conjunction with the new \emph{conjugate\/} fields
$\tilde{\Pi}_{I}^{\mu}$, the $\tilde{R}\indices{_{\eta}^{\alpha\xi\mu}}$
as the duals to $\Gamma\indices{^{\eta}_{\alpha\xi}}(X)$, and the
$\tilde{T}\indices{_{\nu}^{\mu}}$ as the duals to $-\partial X^{\nu}/\partial y^{\mu}$.
In order to derive the pertaining transformation rules,
the following extended Legendre transformation is performed
\begin{eqnarray}
&&\FC_{2}^{\mu}(\phi_{I},\tilde{\Pi}_{I},\gamma,\tilde{R},x,\tilde{T}\,)\nonumber\\
&&\quad=\FC_{1}^{\mu}(\phi_{I},\Phi_{I},\gamma,\Gamma,x,X)+
\Phi_{I}\tilde{\Pi}_{I}^{\mu}\nonumber\\
&&\qquad\mbox{}+\Gamma\indices{^{\eta}_{\alpha\xi}}
\tilde{R}\indices{_{\eta}^{\alpha\xi j}}\pfrac{y^{\mu}}{X^{j}}-X^{j}\tilde{T}\indices{_{j}^{\mu}}.
\label{eq:legendre1-ext}
\end{eqnarray}
The resulting transformation rules are
\begin{eqnarray}
\tilde{\pi}_{I}^{\mu}&=&\pfrac{\FC_{2}^{\mu}}{\phi_{I}},\quad
\Phi_{I}\delta_{\nu}^{\mu}=\pfrac{\FC_{2}^{\mu}}{\tilde{\Pi}_{I}^{\nu}},\quad
\tilde{r}\indices{_{\eta}^{\alpha\xi\mu}}=
\pfrac{\FC_{2}^{j}}{\gamma\indices{^{\eta}_{\alpha\xi}}}\pfrac{x^{\mu}}{y^{j}}\nonumber\\
\Gamma\indices{^{\eta}_{\alpha\xi}}\delta_{\nu}^{\mu}\!&=&\!
\pfrac{\FC_{2}^{\mu}}{\tilde{R}\indices{_{\eta}^{\alpha\xi j}}}\pfrac{X^{j}}{y^{\nu}},\quad\!\!
\tilde{t}\indices{_{\nu}^{\mu}}\!=\!-\pfrac{\FC_{2}^{\mu}}{x^{\nu}},\quad\!\!
X^{\alpha}\delta_{\nu}^{\mu}\!=\!-\pfrac{\FC_{2}^{\mu}}{\tilde{T}\indices{_{\alpha}^{\nu}}}\nonumber\\
\HC_{\e}^{\prime}&=&\HC_{\e},\;
\HC^{\prime}\det\Lambda^{\prime}\!-\!\tilde{T}\indices{_{\alpha}^{\beta}}\pfrac{X^{\alpha}}{y^{\beta}}\!=\!
\HC\det\Lambda\!-\!\tilde{t}\indices{_{\alpha}^{\beta}}\pfrac{x^{\alpha}}{y^{\beta}}.\nonumber\\
\label{eq:genF2-ext}
\end{eqnarray}
These transformation rules are equivalent to the set of Eqs.~(\ref{eq:genF1-ext})
if the Legendre transformation~(\ref{eq:legendre1-ext}) is nonsingular, hence if
the determinant of the Hesse matrix of $\FC_{1}$ is nonzero.
\section{General Space-Time Transformation in Vacuum\label{sec:example-space-time-trans-curv}}
Given a classical vacuum system, i.e., a system with no fields.
An extended canonical transformation that maps the space-time geometry from a space-time
location $x$ to $X$ under the transformation law of the connection coefficients
$\gamma\indices{^{\eta}_{\alpha\xi}}(x)\mapsto\Gamma\indices{^{\eta}_{\alpha\xi}}(X)$
in a \emph{coordinate basis\/} (see Appendix~\ref{app2}) is generated by
\begin{eqnarray}
\FC_{2}^{\mu}(\gamma,\tilde{R},x,\tilde{T}&&\,)=
-\tilde{T}\indices{_{\alpha}^{\mu}}h^{\alpha}(x)
+\tilde{R}\indices{_{\eta}^{\alpha\xi\lambda}}\pfrac{y^{\mu}}{X^{\lambda}}\nonumber\\
&&\times\left(\gamma\indices{^{k}_{ij}}\pfrac{X^{\eta}}{x^{k}}\pfrac{x^{i}}{X^{\alpha}}
\pfrac{x^{j}}{X^{\xi}}+\pfrac{X^{\eta}}{x^{k}}\ppfrac{x^{k}}{X^{\alpha}}{X^{\xi}}\right)\!.\nonumber\\
\label{eq:gen-conn-coeff}
\end{eqnarray}
In this definition of an extended generating function of type $\FC_{2}^{\mu}(y)$,
the tensor density components
$\tilde{R}\indices{_{\eta}^{\alpha\xi\lambda}}(X)\,\partial y^{\mu}/\partial X^{\lambda}$
denote \emph{formally\/} the \emph{canonical conjugates\/} of the connection
coefficients $\Gamma\indices{^{\eta}_{\alpha\xi}}(X)$ of the transformed system
and hence the \emph{dual quantities\/} to the $y^{\mu}$-derivatives of the
$\Gamma\indices{^{\eta}_{\alpha\xi}}(X)$.
The $\gamma\indices{^{\eta}_{\alpha\xi}}(x)$ stand for the connection
coefficients of the original system.
The tensor density components
$\tilde{R}\indices{_{\eta}^{\alpha\xi\mu}}(X)\equiv R\indices{_{\eta}^{\alpha\xi\mu}}(X)\det\Lambda^{\prime}$
then represent the \emph{dual quantities\/} of the $X^{\mu}$-derivatives of the
transformed connection coefficients $\Gamma\indices{^{\eta}_{\alpha\xi}}(X)$
$$
\tilde{R}\indices{_{\eta}^{\alpha\xi\lambda}}(X)\pfrac{y^{\mu}}{X^{\lambda}}
\pfrac{\Gamma\indices{^{\eta}_{\alpha\xi}}(X)}{y^{\mu}}=
R\indices{_{\eta}^{\alpha\xi\lambda}}(X)\pfrac{\Gamma\indices{^{\eta}_{\alpha\xi}}(X)}{X^{\lambda}}
\det\Lambda^{\prime}.
$$
Likewise, the tensor density components
$\tilde{r}\indices{_{\eta}^{\alpha\xi\mu}}(x)\equiv r\indices{_{\eta}^{\alpha\xi\mu}}(x)\det\Lambda$
denote the \emph{dual quantities\/} of the $x^{\mu}$-derivatives of the connection
coefficients $\gamma\indices{^{\eta}_{\alpha\xi}}(x)$ of the original system.
No predication with respect to the physical meaning of
$r\indices{_{\eta}^{\alpha\xi\mu}}$ and $R\indices{_{\eta}^{\alpha\xi\mu}}$ is made at this point.

The particular generating function~(\ref{eq:gen-conn-coeff}) entails the
following transformation rules according to the general rules from Eqs.~(\ref{eq:genF2-ext})
\begin{eqnarray*}
\Gamma\indices{^{\eta}_{\alpha\xi}}\delta_{\nu}^{\mu}&=&
\pfrac{\FC_{2}^{\mu}}{\tilde{R}\indices{_{\eta}^{\alpha\xi j}}}\pfrac{X^{j}}{y^{\nu}}\nonumber\\
&=&\delta_{\nu}^{\mu}
\left(\gamma\indices{^{k}_{ij}}\pfrac{X^{\eta}}{x^{k}}\pfrac{x^{i}}{X^{\alpha}}\pfrac{x^{j}}{X^{\xi}}+
\pfrac{X^{\eta}}{x^{k}}\ppfrac{x^{k}}{X^{\alpha}}{X^{\xi}}\right)\\
\tilde{r}\indices{_{k}^{ij\mu}}&=&
\pfrac{\FC_{2}^{\kappa}}{\gamma\indices{^{k}_{ij}}}\pfrac{x^{\mu}}{y^{\kappa}}=
\tilde{R}\indices{_{\eta}^{\alpha\xi\lambda}}\pfrac{X^{\eta}}{x^{k}}
\pfrac{x^{i}}{X^{\alpha}}\pfrac{x^{j}}{X^{\xi}}\pfrac{x^{\mu}}{X^{\lambda}}\\
X^{\alpha}\delta_{\nu}^{\mu}&=&-\pfrac{\FC_{2}^{\mu}}{\tilde{T}\indices{_{\alpha}^{\nu}}}=
\delta_{\nu}^{\mu}h^{\alpha}(x)\\
\tilde{t}\indices{_{\nu}^{\mu}}&=&-\pfrac{\FC_{2}^{\mu}}{x^{\nu}}\\
&=&\tilde{T}\indices{_{\alpha}^{\mu}}\pfrac{h^{\alpha}(x)}{x^{\nu}}-
\tilde{R}\indices{_{\eta}^{\alpha\xi\lambda}}\pfrac{y^{\mu}}{X^{\lambda}}\\
&&\times\left[\gamma\indices{^{k}_{ij}}\pfrac{}{x^{\nu}}\left(
\pfrac{X^{\eta}}{x^{k}}\pfrac{x^{i}}{X^{\alpha}}\pfrac{x^{j}}{X^{\xi}}\right)\right.\\
&&\qquad\left.\mbox{}+\pfrac{}{x^{\nu}}\left(\pfrac{X^{\eta}}{x^{k}}\ppfrac{x^{k}}{X^{\alpha}}{X^{\xi}}\right)\right]
\end{eqnarray*}
hence
\begin{eqnarray}
\Gamma\indices{^{\eta}_{\alpha\xi}}(X)&=&
\gamma\indices{^{k}_{ij}}(x)\pfrac{X^{\eta}}{x^{k}}\pfrac{x^{i}}{X^{\alpha}}\pfrac{x^{j}}{X^{\xi}}+
\pfrac{X^{\eta}}{x^{k}}\ppfrac{x^{k}}{X^{\alpha}}{X^{\xi}}\nonumber\\
\tilde{r}\indices{_{k}^{ij\mu}}(x)&=&
\tilde{R}\indices{_{\eta}^{\alpha\xi\lambda}}(X)\pfrac{X^{\eta}}{x^{k}}
\pfrac{x^{i}}{X^{\alpha}}\pfrac{x^{j}}{X^{\xi}}\pfrac{x^{\mu}}{X^{\lambda}}\nonumber\\
X^{\mu}&=&h^{\mu}(x)\nonumber\\
\tilde{t}\indices{_{\nu}^{\mu}}&=&
\tilde{T}\indices{_{\alpha}^{\mu}}\pfrac{X^{\alpha}}{x^{\nu}}-
\tilde{R}\indices{_{\eta}^{\alpha\xi\lambda}}\pfrac{y^{\mu}}{X^{\lambda}}\nonumber\\
&&\times\left[\gamma\indices{^{k}_{ij}}\pfrac{}{x^{\nu}}\left(
\pfrac{X^{\eta}}{x^{k}}\pfrac{x^{i}}{X^{\alpha}}\pfrac{x^{j}}{X^{\xi}}\right)\right.\nonumber\\
&&\qquad\left.\mbox{}+\pfrac{}{x^{\nu}}\left(\pfrac{X^{\eta}}{x^{k}}\ppfrac{x^{k}}{X^{\alpha}}{X^{\xi}}\right)\right].
\label{eq:tr-gen-space-time}
\end{eqnarray}
According to Eqs.~(\ref{eq:genF2-ext}), the last rule yields
the transformation rule for the conventional Hamiltonians via
\begin{eqnarray}
&&\tilde{T}\indices{_{\alpha}^{\beta}}\pfrac{X^{\alpha}}{y^{\beta}}-
\tilde{t}\indices{_{\alpha}^{\beta}}\pfrac{x^{\alpha}}{y^{\beta}}\nonumber\\
&&\quad=\HC^{\prime}\det\Lambda^{\prime}-\HC\det\Lambda\nonumber\\
&&\quad=\tilde{R}\indices{_{\eta}^{\alpha\xi\lambda}}\pfrac{x^{\nu}}{X^{\lambda}}
\left[\gamma\indices{^{k}_{ij}}\pfrac{}{x^{\nu}}\left(
\pfrac{X^{\eta}}{x^{k}}\pfrac{x^{i}}{X^{\alpha}}\pfrac{x^{j}}{X^{\xi}}\right)\right.\nonumber\\
&&\qquad\qquad\qquad\qquad\left.\mbox{}+\pfrac{}{x^{\nu}}
\left(\pfrac{X^{\eta}}{x^{k}}\ppfrac{x^{k}}{X^{\alpha}}{X^{\xi}}\right)\right].
\label{eq:ham-conn-coeff}
\end{eqnarray}
Yet, what is actually desired is the transformation rule for the Hamiltonians as expressed
in terms of their proper dynamical variables $\gamma,r$ and $\Gamma,R$, respectively.
This requires to express all derivatives of the functions $x^{\mu}(y)$ and $X^{\mu}(y)$
in~(\ref{eq:ham-conn-coeff}) in terms of the original and transformed connection
coefficients $\gamma\indices{^{\eta}_{\alpha\xi}}(x)$ and
$\Gamma\indices{^{\eta}_{\alpha\xi}}(X)$ and their conjugates,
$\tilde{r}\indices{_{\eta}^{\alpha\xi\mu}}$ and $\tilde{R}\indices{_{\eta}^{\alpha\xi\mu}}$,
by making use of the respective canonical transformation rules~(\ref{eq:tr-gen-space-time}).
This calculation is worked out explicitly in Appendix~\ref{app4}.
Remarkably, the transformation rule~(\ref{eq:ham-conn-coeff}) can indeed
\emph{completely and symmetrically\/} be expressed in terms of the canonical
variables of the original and the transformed system as
\begin{eqnarray*}
&&\tilde{T}\indices{_{\alpha}^{\beta}}\pfrac{X^{\alpha}}{y^{\beta}}-
\tilde{t}\indices{_{\alpha}^{\beta}}\pfrac{x^{\alpha}}{y^{\beta}}\\
&&\quad=\onehalf\tilde{R}\indices{_{\eta}^{\alpha\xi\mu}}\!\left(\!
\pfrac{\Gamma\indices{^{\eta}_{\alpha\xi}}}{X^{\mu}}+\pfrac{\Gamma\indices{^{\eta}_{\alpha\mu}}}{X^{\xi}}-
\Gamma\indices{^{k}_{\alpha\xi}}\Gamma\indices{^{\eta}_{k\mu}}+
\Gamma\indices{^{k}_{\alpha\mu}}\Gamma\indices{^{\eta}_{k\xi}}\right)\\
&&\qquad\mbox{}-\onehalf\tilde{r}\indices{_{\eta}^{\alpha\xi\mu}}\!\!\left(\!
\pfrac{\gamma\indices{^{\eta}_{\alpha\xi}}}{x^{\mu}}\!+\!\pfrac{\gamma\indices{^{\eta}_{\alpha\mu}}}{x^{\xi}}-
\gamma\indices{^{k}_{\alpha\xi}}\gamma\indices{^{\eta}_{k\mu}}\!+\!
\gamma\indices{^{k}_{\alpha\mu}}\gamma\indices{^{\eta}_{k\xi}}\!\right)\!.
\end{eqnarray*}
Gathering original and transformed dynamical variables on either side of the equation yields
\begin{eqnarray}
&&\onehalf\tilde{R}\indices{_{\eta}^{\alpha\xi\mu}}\left(
\pfrac{\Gamma\indices{^{\eta}_{\alpha\xi}}}{X^{\mu}}+\pfrac{\Gamma\indices{^{\eta}_{\alpha\mu}}}{X^{\xi}}-
\Gamma\indices{^{k}_{\alpha\xi}}\Gamma\indices{^{\eta}_{k\mu}}+
\Gamma\indices{^{k}_{\alpha\mu}}\Gamma\indices{^{\eta}_{k\xi}}\right)\nonumber\\
&&\qquad\mbox{}-\tilde{T}\indices{_{\alpha}^{\beta}}\pfrac{X^{\alpha}}{y^{\beta}}\nonumber\\
&&=\onehalf\tilde{r}\indices{_{\eta}^{\alpha\xi\mu}}\left(
\pfrac{\gamma\indices{^{\eta}_{\alpha\xi}}}{x^{\mu}}+\pfrac{\gamma\indices{^{\eta}_{\alpha\mu}}}{x^{\xi}}-
\gamma\indices{^{k}_{\alpha\xi}}\gamma\indices{^{\eta}_{k\mu}}+
\gamma\indices{^{k}_{\alpha\mu}}\gamma\indices{^{\eta}_{k\xi}}\right)\nonumber\\
&&\qquad\mbox{}-\tilde{t}\indices{_{\alpha}^{\beta}}\pfrac{x^{\alpha}}{y^{\beta}}.
\label{eq:ham-conn-coeff1}
\end{eqnarray}
The left- and right-hand sides of this equation can be regarded as
extended Hamiltonians $\HC_{\e}^{\prime}$ and $\HC_{\e}$, respectively,
which satisfy the required transformation rule $\HC_{\e}^{\prime}=\HC_{\e}$
from Eqs.~(\ref{eq:genF2-ext}).
Obviously, the Hamiltonians not only retain their values but are furthermore
\emph{form-invariant\/} under the extended canonical transformation generated
by Eq.~(\ref{eq:gen-conn-coeff}).

In order for the canonical equations following from the Hamiltonians $\HC_{\e}$
and $\HC_{\e}^{\prime}$ to be compatible with the
canonical transformation rules~(\ref{eq:tr-gen-space-time}), the above
form-invariant Hamiltonians must be amended, by ``free field'' Hamiltonians
\begin{eqnarray*}
\HC^{\prime}_{\e,\mathrm{dyn}}&=&-\quarter
R\indices{_{\eta}^{\alpha\xi\mu}}R\indices{^{\eta}_{\alpha\xi\mu}}\det\Lambda^{\prime},\\
\HC_{\e,\mathrm{dyn}}&=&-\quarter r\indices{_{\eta}^{\alpha\xi\mu}}r\indices{^{\eta}_{\alpha\xi\mu}}\det\Lambda.
\end{eqnarray*}
Clearly, $\HC^{\prime}_{\e,\mathrm{dyn}}=\HC_{\e,\mathrm{dyn}}$ must hold
under the rules~(\ref{eq:tr-gen-space-time}) in order for the final extended
Hamiltonians to maintain the required transformation rule $\HC_{\e}^{\prime}=\HC_{\e}$.
This is ensured if $\det\Lambda^{\prime}=\det\Lambda$, hence if
\begin{equation}\label{eq:h-alpha-cond}
\pfrac{\left(X^{0},\ldots,X^{3}\right)}{\left(x^{0},\ldots,x^{3}\right)}=
\pfrac{\left(h^{0}(x),\ldots,h^{3}(x)\right)}{\left(x^{0},\ldots,x^{3}\right)}=1.
\end{equation}
Thus, the by now arbitrary function $h^{\alpha}(x)$ in the generating function~(\ref{eq:gen-conn-coeff})
must satisfy Eq.~(\ref{eq:h-alpha-cond}).
The final form-invariant extended Hamiltonian that is compatible with the extended set of canonical
transformation rules generated by~(\ref{eq:gen-conn-coeff}) now writes for the $x$ reference frame
\begin{eqnarray}
\HC_{\e}&&\left(\tilde{r},\gamma,\tilde{t}\,\right)=
-\tilde{t}\indices{_{\alpha}^{\beta}}\pfrac{x^{\alpha}}{y^{\beta}}-
\quarter\tilde{r}\indices{_{\eta}^{\alpha\xi\mu}}\,r\indices{^{\eta}_{\alpha\xi\mu}}\nonumber\\
&&\;\mbox{}+\onehalf\tilde{r}\indices{_{\eta}^{\alpha\xi\mu}}\!\left(\!
\pfrac{\gamma\indices{^{\eta}_{\alpha\mu}}}{x^{\xi}}+\pfrac{\gamma\indices{^{\eta}_{\alpha\xi}}}{x^{\mu}}+
\gamma\indices{^{k}_{\alpha\mu}}\gamma\indices{^{\eta}_{k\xi}}-
\gamma\indices{^{k}_{\alpha\xi}}\gamma\indices{^{\eta}_{k\mu}}\!\right)\!.\nonumber\\
\label{eq:eham-conn-coeff-inv}
\end{eqnarray}
\section{Canonical equations}
The canonical equation for the connection coefficients follows as
\begin{eqnarray*}
\pfrac{\gamma\indices{^{\eta}_{\alpha\xi}}}{x^{\mu}}&=&
\pfrac{\HC_{\e}}{\tilde{r}\indices{_{\eta}^{\alpha\xi\mu}}}\\
&=&-\onehalf r\indices{^{\eta}_{\alpha\xi\mu}}\\
&&\!\mbox{}+\onehalf\!\left(\!
\pfrac{\gamma\indices{^{\eta}_{\alpha\mu}}}{x^{\xi}}+
\pfrac{\gamma\indices{^{\eta}_{\alpha\xi}}}{x^{\mu}}+
\gamma\indices{^{k}_{\alpha\mu}}\gamma\indices{^{\eta}_{k\xi}}-
\gamma\indices{^{k}_{\alpha\xi}}\gamma\indices{^{\eta}_{k\mu}}\!\right)\!.
\end{eqnarray*}
Solved for $r\indices{^{\eta}_{\alpha\xi\mu}}$ one finds
\begin{equation}\label{eq:Riemann-tensor2}
r\indices{^{\eta}_{\alpha\xi\mu}}=\pfrac{\gamma\indices{^{\eta}_{\alpha\mu}}}{x^{\xi}}-
\pfrac{\gamma\indices{^{\eta}_{\alpha\xi}}}{x^{\mu}}+
\gamma\indices{^{k}_{\alpha\mu}}\gamma\indices{^{\eta}_{k\xi}}-
\gamma\indices{^{k}_{\alpha\xi}}\gamma\indices{^{\eta}_{k\mu}}.
\end{equation}
One thus encounters exactly the representation of the Riemann curvature tensor
in terms of the connection coefficients and their derivatives.
The quantity $\tilde{R}\indices{_{\eta}^{\alpha\xi\lambda}}(X)\partial y^{\mu}/\partial X^{\lambda}$
was formally introduced in the generating function~(\ref{eq:gen-conn-coeff}) as the
canonical conjugate of the connection coefficient $\Gamma\indices{^{\eta}_{\alpha\xi}}(X)$
in order to yield the required transformation law for $\gamma\indices{^{\eta}_{\alpha\xi}}(x)$.
With Eq.~(\ref{eq:Riemann-tensor2}), the quantity $r\indices{_{\eta}^{\alpha\xi\mu}}$
is \emph{now\/} attributed a physical meaning.
It is manifestly skew-symmetric in the indices $\xi$ and $\mu$.
With Eq.~(\ref{eq:Riemann-tensor2}) being a \emph{tensor equation},
the second canonical transformation~(\ref{eq:tr-gen-space-time}) rule is satisfied,
which requires $r\indices{^{\eta}_{\alpha\xi\mu}}$ to transform as a \emph{tensor}.

The divergence of $\tilde{r}\indices{_{\eta}^{\alpha\xi\mu}}$ follows from the
derivative of $\HC_{\e}$ with respect to $\gamma\indices{^{\eta}_{\alpha\xi}}$
\begin{equation}\label{eq:feq-conn-coeff}
{\left(\tilde{r}\indices{_{\eta}^{\alpha\xi\beta}}\right)}_{,\beta}\equiv
\pfrac{\tilde{r}\indices{_{\eta}^{\alpha\xi\beta}}}{x^{\beta}}=
-\pfrac{\HC_{\e}}{\gamma\indices{^{\eta}_{\alpha\xi}}}=
\gamma\indices{^{k}_{\eta\beta}}\tilde{r}\indices{_{k}^{\alpha\xi\beta}}-
\gamma\indices{^{\alpha}_{k\beta}}\tilde{r}\indices{_{\eta}^{k\xi\beta}}.
\end{equation}
On the other hand, the \emph{covariant\/} divergence of a tensor density
$\tilde{r}\indices{_{\eta}^{\alpha\xi\mu}}$ is given by
\begin{eqnarray*}
{\left(\tilde{r}\indices{_{\eta}^{\alpha\xi\beta}}\right)}_{;\beta}&=&
{\left(\tilde{r}\indices{_{\eta}^{\alpha\xi\beta}}\right)}_{,\beta}-
\gamma\indices{^{k}_{\eta\beta}}\tilde{r}\indices{_{k}^{\alpha\xi\beta}}+
\gamma\indices{^{\alpha}_{\beta k}}\tilde{r}\indices{_{\eta}^{k\xi\beta}}\\
&&\mbox{}+
\gamma\indices{^{\xi}_{\beta k}}\tilde{r}\indices{_{\eta}^{\alpha k\beta}}+
\cancel{\gamma\indices{^{\beta}_{\beta k}}\tilde{r}\indices{_{\eta}^{\alpha\xi k}}}-
\cancel{\gamma\indices{^{k}_{k\beta}}\tilde{r}\indices{_{\eta}^{\alpha\xi\beta}}}.
\end{eqnarray*}
The last two terms cancel as usual for the divergence of a tensor density.
The field equation~(\ref{eq:feq-conn-coeff}) is thus equivalently expressed
in terms of the covariant derivative as
\begin{eqnarray}
{\left(\tilde{r}\indices{_{\eta}^{\alpha\xi\beta}}\right)}_{;\beta}&=&
\gamma\indices{^{\alpha}_{k\beta}}\tilde{r}\indices{_{\eta}^{k\beta\xi}}+
\gamma\indices{^{\alpha}_{\beta k}}\tilde{r}\indices{_{\eta}^{k\xi\beta}}+
\gamma\indices{^{\xi}_{\beta k}}\tilde{r}\indices{_{\eta}^{\alpha k\beta}}\nonumber\\
&=&\left(\gamma\indices{^{\alpha}_{k\beta}}-\gamma\indices{^{\alpha}_{\beta k}}\right)
\tilde{r}\indices{_{\eta}^{k\beta\xi}}-
\gamma\indices{^{\xi}_{k\beta}}\tilde{r}\indices{_{\eta}^{\alpha k\beta}}.
\label{eq:feq-conn-coeff4}
\end{eqnarray}
As $\tilde{r}\indices{_{\eta}^{\alpha k\beta}}$ is skew-symmetric in $k,\beta$ according to
the first canonical equation~(\ref{eq:Riemann-tensor2}), the contraction with
$\gamma\indices{^{\xi}_{k\beta}}$ in the rightmost term of Eq.~(\ref{eq:feq-conn-coeff4})
extracts the skew-symmetric part of the connection coefficients, hence the \emph{torsion tensor\/}
$s\indices{^{\xi}_{k\beta}}=\onehalf(\gamma\indices{^{\xi}_{k\beta}}-\gamma\indices{^{\xi}_{\beta k}})$.
Thus, Eq.~(\ref{eq:feq-conn-coeff4}) is actually a \emph{tensor equation}
\begin{eqnarray}
&&{\left(\tilde{r}\indices{_{\eta}^{\alpha\xi\beta}}\right)}_{;\beta}\nonumber\\
&&=\cases{
2s\indices{^{\alpha}_{k\beta}}\tilde{r}\indices{_{\eta}^{k\beta\xi}}-
s\indices{^{\xi}_{k\beta}}\tilde{r}\indices{_{\eta}^{\alpha k\beta}}\!\!\!&in general\cr
0&for a torsion-free space-time.}\nonumber\\
\label{eq:feq-conn-coeff3}
\end{eqnarray}
The canonical equation for the space-time coefficients follows as
\begin{displaymath}
\pfrac{x^{\nu}}{y^{\mu}}=
-\pfrac{\HC_{\e}}{\tilde{t}\indices{_{\nu}^{\mu}}}=\pfrac{x^{\nu}}{y^{\mu}}.
\end{displaymath}
As a common feature of all trivial extended Hamiltonians, no
\emph{substantial\/} equation for the space-time coefficients
emerges but only an identity that allows for arbitrary space-time dynamics.

As $\HC_{\e}$ does not depend on the $x^{\nu}$,
the canonical equation for the $\tilde{t}\indices{_{\nu}^{\mu}}$ follows as
\begin{equation}\label{eq:caneq-QG}
\pfrac{\tilde{t}\indices{_{\nu}^{\alpha}}}{y^{\alpha}}=
-\pfrac{\HC_{\e}}{x^{\nu}}=0.
\end{equation}
The explicit representation of $\tilde{t}\indices{_{\nu}^{\alpha}}$ will
be derived in Eq.~(\ref{eq:pext-QG}) from the Lagrangian $\LC_{\e}$ that
follows from $\HC_{\e}$ by means of the Legendre transformation prescription~(\ref{H1-def1}).
\section{Form-invariant Lagrangian, Euler-Lagrange equations}
The extended Lagrangian $\LC_{\e}$ corresponding to the form-invariant
Hamiltonian from~(\ref{eq:eham-conn-coeff-inv}) is obtained by means of
the regular Legendre transformation
\begin{eqnarray}
\LC_{\e}&=&\tilde{r}\indices{_{\eta}^{\alpha\xi\mu}}
\pfrac{\gamma\indices{^{\eta}_{\alpha\xi}}}{x^{\mu}}-
\tilde{t}\indices{_{\alpha}^{\beta}}\pfrac{x^{\alpha}}{y^{\beta}}-\HC_{\e}\nonumber\\
&=&\quarter\tilde{r}\indices{_{\eta}^{\alpha\xi\mu}}\,r\indices{^{\eta}_{\alpha\xi\mu}}
-\onehalf\tilde{r}\indices{_{\eta}^{\alpha\xi\mu}}\nonumber\\
&&\times\left(
\pfrac{\gamma\indices{^{\eta}_{\alpha\mu}}}{x^{\xi}}-\pfrac{\gamma\indices{^{\eta}_{\alpha\xi}}}{x^{\mu}}+
\gamma\indices{^{k}_{\alpha\mu}}\gamma\indices{^{\eta}_{k\xi}}-
\gamma\indices{^{k}_{\alpha\xi}}\gamma\indices{^{\eta}_{k\mu}}\right)\nonumber\\
&=&-\quarter\tilde{r}\indices{_{\eta}^{\alpha\xi\mu}}\,r\indices{^{\eta}_{\alpha\xi\mu}}.
\label{eq:Lag-conn-coeff}
\end{eqnarray}
Note that---in contrast to the Hamiltonian description---the curvature tensor
$r\indices{^{\eta}_{\alpha\xi\mu}}$ does not constitute a proper dynamical variable
but only abbreviates the particular combination~(\ref{eq:Riemann-tensor2})
of the connection coefficients and their respective derivatives---which
comprise in conjunction with the space-time coefficients
the actual dynamical variables of the Lagrangian description.
The dependence of $\LC_{\e}$ on the space-time coefficients
is expressed implicitly in terms of metric tensors
\begin{eqnarray}
\LC_{\e}&&\left(\gamma\indices{^{\eta}_{\alpha\xi}},\pfrac{\gamma\indices{^{\eta}_{\alpha\xi}}}{x^{\nu}},
\pfrac{x^{\mu}}{y^{\nu}}\right)=
-\quarter r\indices{_{\eta}^{\alpha\xi\mu}}\,r\indices{^{\eta}_{\alpha\xi\mu}}\det\Lambda\nonumber\\
&&\qquad=-\quarter g_{\kappa\eta}g^{\beta\alpha}g^{\lambda\xi}g^{\zeta\mu}r\indices{^{\kappa}_{\beta\lambda\zeta}}
r\indices{^{\eta}_{\alpha\xi\mu}}\!\det\Lambda.
\label{eq:Lag-conn-coeff2}
\end{eqnarray}
The $\tilde{t}\indices{_{\mu}^{\nu}}$ represent the \emph{duals\/}
of the space-time coefficients, hence, their explicit form follows
from $\LC_{\e}$ according to the general rule~(\ref{pext-def}).
Owing to the identities~(\ref{eq:deri-cov-metric3}) and~(\ref{eq:deri-cov-metric4}),
one finds for the Lagrangian~(\ref{eq:Lag-conn-coeff2})
\begin{eqnarray}
-\tilde{t}\indices{_{\mu}^{\nu}}&=&\pfrac{\LC_{\e}}{\left(\pfrac{x^{\mu}}{y^{\nu}}\right)}\nonumber\\
&=&\!\left(\!r\indices{_{\eta}^{\alpha\xi k}}(x)\,r\indices{^{\eta}_{\alpha\xi\mu}}(x)\!-\!
\quarter\delta_{\mu}^{k}\,r\indices{_{\eta}^{\alpha\xi\beta}}\,r\indices{^{\eta}_{\alpha\xi\beta}}\!
\right)\!\pfrac{y^{\nu}}{x^{k}}\det\Lambda.\nonumber\\
\label{eq:pext-QG}
\end{eqnarray}
The explicit calculation is worked out in Appendix~\ref{app5}.
As $\LC_{\e}$ does not depend on $x^{\mu}$, the divergence of $\tilde{t}\indices{_{\mu}^{\nu}}$ vanishes
according to the Euler-Lagrange equation~(\ref{elgl-ext1}) and in agreement with the
canonical equation~(\ref{eq:caneq-QG}).
With Eq.~(\ref{eq:ext-lag-identity}), its final form is then
\begin{displaymath}
\pfrac{}{x^{k}}\left(r\indices{_{\eta}^{\alpha\xi k}}(x)\,r\indices{^{\eta}_{\alpha\xi\mu}}(x)-
\quarter\delta_{\mu}^{k}\,r\indices{_{\eta}^{\alpha\xi\beta}}\,
r\indices{^{\eta}_{\alpha\xi\beta}}\right)=0,
\end{displaymath}
which has the corresponding tensor form
\begin{equation}\label{eq:sec-order-einstein0}
{\left(r\indices{_{\eta}^{\alpha\xi k}}\,r\indices{^{\eta}_{\alpha\xi\mu}}-
\quarter\delta_{\mu}^{k}\,r\indices{_{\eta}^{\alpha\xi\beta}}\,
r\indices{^{\eta}_{\alpha\xi\beta}}\right)}_{;k}=0.
\end{equation}
To set up the Euler-Lagrange equation~(\ref{elgl-ext2}) for the connection coefficients,
the derivative of $\LC_{\e}$ from Eq.~(\ref{eq:Lag-conn-coeff2}) with respect to the derivatives
of the connection coefficients is to be calculated first
\begin{eqnarray}
\pfrac{\LC_{\e}}{\left(\pfrac{\gamma\indices{^{k}_{ij}}(x)}{x^{\lambda}}\right)}&=&
-\onehalf\tilde{r}\indices{_{\eta}^{\alpha\xi\mu}}
\pfrac{r\indices{^{\eta}_{\alpha\xi\mu}}}{\left(\pfrac{\gamma\indices{^{k}_{ij}}}{x^{\lambda}}\right)}\nonumber\\
&=&-\onehalf\tilde{r}\indices{_{\eta}^{\alpha\xi\mu}}\delta_{k}^{\eta}\left(
\delta_{\alpha}^{i}\delta_{\xi}^{\lambda}\delta_{\mu}^{j}-
\delta_{\alpha}^{i}\delta_{\xi}^{j}\delta_{\mu}^{\lambda}\right)\nonumber\\
&=&-\onehalf\left(\tilde{r}\indices{_{k}^{i\lambda j}}-\tilde{r}\indices{_{k}^{ij\lambda}}\right)\nonumber\\
&=&\tilde{r}\indices{_{k}^{ij\lambda}}(x).
\label{eq:gamma-rtilde-duality}
\end{eqnarray}
The quantities $\partial\gamma\indices{^{k}_{ij}}(x)/\partial x^{\lambda}$ and $\tilde{r}\indices{_{k}^{ij\lambda}}(x)$
are thus dual to each other in the system described by the extended Lagrangian~(\ref{eq:Lag-conn-coeff2}),
in agreement with the corresponding canonical equation~(\ref{eq:Riemann-tensor2}).

The derivative of $\LC_{\e}$ with respect to the connection coefficients is
\begin{eqnarray*}
\pfrac{\LC_{\e}}{\gamma\indices{^{k}_{ij}}}&=&
-\onehalf\tilde{r}\indices{_{\eta}^{\alpha\beta\mu}}\left(
\delta_{k}^{\xi}\delta_{\alpha}^{i}\delta_{\mu}^{j}\,\gamma\indices{^{\eta}_{\xi\beta}}+
\delta_{\xi}^{i}\delta_{\beta}^{j}\delta_{k}^{\eta}\,\gamma\indices{^{\xi}_{\alpha\mu}}\right.\\
&&\qquad\qquad\quad\left.\mbox{}-\delta_{k}^{\xi}\delta_{\alpha}^{i}\delta_{\beta}^{j}\,\gamma\indices{^{\eta}_{\xi\mu}}-
\delta_{\xi}^{i}\delta_{\mu}^{j}\delta_{k}^{\eta}\,\gamma\indices{^{\xi}_{\alpha\beta}}\right)\\
&=&-\onehalf\left(
\tilde{r}\indices{_{\eta}^{i\beta j}}\gamma\indices{^{\eta}_{k\beta}}+
\tilde{r}\indices{_{k}^{\alpha j\mu}}\gamma\indices{^{i}_{\alpha\mu}}\right.\\
&&\qquad\left.\mbox{}-
\tilde{r}\indices{_{\eta}^{ij\mu}}\gamma\indices{^{\eta}_{k\mu}}-
\tilde{r}\indices{_{k}^{\alpha\beta j}}\gamma\indices{^{i}_{\alpha\beta}}\right)\\
&=&\tilde{r}\indices{_{\beta}^{ij\alpha}}\gamma\indices{^{\beta}_{k\alpha}}+
\tilde{r}\indices{_{k}^{\alpha\beta j}}\gamma\indices{^{i}_{\alpha\beta}}.
\end{eqnarray*}
In conjunction with Eq.~(\ref{eq:gamma-rtilde-duality}) this yields the Euler-Lagrange equation
\begin{displaymath}
\pfrac{\tilde{r}\indices{_{k}^{ij\lambda}}}{x^{\lambda}}-
\gamma\indices{^{\beta}_{k\alpha}}\tilde{r}\indices{_{\beta}^{ij\alpha}}-
\gamma\indices{^{i}_{\alpha\beta}}\tilde{r}\indices{_{k}^{\alpha\beta j}}=0,
\end{displaymath}
which actually represents a \emph{tensor equation\/} and
agrees, as expected, with the canonical equation from Eq.~(\ref{eq:feq-conn-coeff})
and the subsequent field equations~(\ref{eq:feq-conn-coeff3}).
For a torsion-free space-time, one gets in particular
\begin{equation}\label{eq:sec-order-einstein2}
{\left(r\indices{_{\eta}^{\alpha\xi k}}\right)}_{;k}=0,
\end{equation}
which is a sufficient condition for Eqs.~(\ref{eq:sec-order-einstein0})
to be satisfied \emph{identically\/}~\cite{stephenson58}.

The coupled set of field equations~(\ref{eq:sec-order-einstein0}) and~(\ref{eq:sec-order-einstein2})
must be simultaneously satisfied in order to minimize the action~(\ref{action-int3}).
Equation~(\ref{eq:sec-order-einstein2}) provides the correlation of the connection coefficients with the metric,
which here does not coincide with the usual Levi-Civita connection of standard general relativity.
\section{Form-invariant Lagrangian including matter}
The canonical approach to general relativity
suggests that the dynamics of space-time may be described by an extended
Lagrangian scalar density that is \emph{quadratic\/} in the Riemann curvature tensor.
Explicitly, this ``quadratic gravity'' Lagrangian is proposed as
\begin{equation}\label{eq:higher-curvature-lag}
\LC_{\e}^{\mathrm{QG}}=\LC_{\e}+\bar{\kappa}\LC_{\mathrm{M}}\det\Lambda,\qquad
\LC_{\e}=-\quarter r\indices{_{\eta}^{\alpha\xi\beta}}r\indices{^{\eta}_{\alpha\xi\beta}}\det\Lambda,
\end{equation}
with $\bar{\kappa}$ a dimensionless coupling constant to the subsystem $\LC_{\mathrm{M}}$ that
describes a conventional dynamical system associated with mass and/or energy.
Therefore $\LC_{\mathrm{M}}\det\Lambda$ defines a \emph{trivial\/} extended Lagrangian.
Its derivative with respect to the space-time coefficients then yields
the canonical energy-momentum tensor $\theta\indices{_{\mu}^{\nu}}(x)$ of the
system described by $\LC_{\mathrm{M}}$, as derived in Eq.~(\ref{L1-deri}).
The derivative of $\LC_{\e}^{\mathrm{QG}}$ then follows as
\begin{eqnarray*}
\pfrac{\LC_{\e}^{\mathrm{QG}}}{\left(\pfrac{x^{\mu}}{y^{\nu}}\right)}&=&
\Big[r\indices{_{\eta}^{\alpha\xi k}}(x)\,r\indices{^{\eta}_{\alpha\xi\mu}}(x)-
\quarter\delta_{\mu}^{k}\,r\indices{_{\eta}^{\alpha\xi\beta}}\,r\indices{^{\eta}_{\alpha\xi\beta}}\\
&&\quad\mbox{}-\bar{\kappa}\theta\indices{_{\mu}^{k}}(x)\Big]\pfrac{y^{\nu}}{x^{k}}\det\Lambda.
\end{eqnarray*}
If $\LC_{\mathrm{M}}$ does not explicitly depend on the $x^{\mu}$, then the Euler-Lagrange
equation for the space-time coefficients is given by
\begin{eqnarray*}
\pfrac{}{y^{\nu}}&&\left(
\Big[r\indices{_{\eta}^{\alpha\xi k}}(x)\,r\indices{^{\eta}_{\alpha\xi\mu}}(x)-
\quarter\delta_{\mu}^{k}\,r\indices{_{\eta}^{\alpha\xi\beta}}\,r\indices{^{\eta}_{\alpha\xi\beta}}\right.\\
&&\left.\quad\mbox{}-\bar{\kappa}\theta\indices{_{\mu}^{k}}(x)\Big]\pfrac{y^{\nu}}{x^{k}}\det\Lambda\right)=0,
\end{eqnarray*}
hence by virtue of Eq.~(\ref{eq:ext-lag-identity}) in tensor form
\begin{equation}\label{eq:sec-order-einstein}
{\Big(r\indices{_{\eta}^{\alpha\xi k}}\,r\indices{^{\eta}_{\alpha\xi\mu}}-
\quarter\delta_{\mu}^{k}\,r\indices{_{\eta}^{\alpha\xi\beta}}\,
r\indices{^{\eta}_{\alpha\xi\beta}}-\bar{\kappa}\theta\indices{_{\mu}^{k}}\Big)}_{;k}=0,
\end{equation}
which generalizes the field equation of the matter-free system from Eq.~(\ref{eq:sec-order-einstein0}).
As $\LC_{\mathrm{M}}\det\Lambda$ does not depend on the connection coefficients,
the corresponding Euler-Lagrange equation of $\LC_{\e}^{\mathrm{QG}}$
agrees with~(\ref{eq:sec-order-einstein2}).
Provided that the conventional mass/energy Lagrangian $\LC_{\mathrm{M}}$ in
(\ref{eq:higher-curvature-lag}) is \emph{autonomous}, hence does not explicitly depend on space-time,
then both groups of covariant derivatives of (\ref{eq:sec-order-einstein}) vanish separately
\begin{displaymath}
{\Big(r\indices{_{\eta}^{\alpha\xi k}}\,r\indices{^{\eta}_{\alpha\xi\mu}}-
\quarter\delta_{\mu}^{k}\,r\indices{_{\eta}^{\alpha\xi\beta}}\,
r\indices{^{\eta}_{\alpha\xi\beta}}\Big)}_{;k}\equiv0,\qquad\theta\indices{_{\mu}^{k}_{;k}}\equiv0,
\end{displaymath}
which states the local conservation of energy and momentum for a closed system.

Equation~(\ref{eq:sec-order-einstein}) constitutes the equation of general relativity
as derived from the extended canonical transformation of the connection coefficients.
It is \emph{not\/} assumed here that the Levi-Civita connection applies for
the independent geometric quantities given by the Christoffel symbols and the metric.
Rather, the connection is established by Eq.~(\ref{eq:feq-conn-coeff3})
or by the more special equation~(\ref{eq:sec-order-einstein2}) in the
case of a torsion-free space-time.
\section{Conclusions}
The gauge theory that is based on the extended canonical transformation
formalism suggests that general relativity is described by the Lagrangian
\begin{equation}\label{eq:lag-genrel}
\LC_{\e}^{\mathrm{QG}}=\left(-\quarter R\indices{_{\eta}^{\alpha\xi\beta}}R\indices{^{\eta}_{\alpha\xi\beta}}+
\bar{\kappa}\LC_{\mathrm{M}}\right)\det\Lambda,
\end{equation}
with $\bar{\kappa}$ a dimensionless coupling constant of the Riemann curvature
tensor term and $\LC_{\mathrm{M}}$ the conventional matter/energy Lagrangian.
Following Palatini's approach, the extended canonical formalism treats the Christoffel symbols
and the space-time coefficients implicitly contained in~(\ref{eq:lag-genrel}) as \emph{independent quantities}.
This yields two independent Euler-Lagrange equations that must be \emph{simultaneously\/}
satisfied in order for the solutions to satisfy the action principle.
No \emph{a priori\/} assumptions on a correlation of the Christoffel symbols
with the space-time coefficients---and hence the metric---are made.

The extended Lagrangian~(\ref{eq:lag-genrel}) was obtained by Legendre-transforming
the form-invariant extended Hamiltonian that emerged from an extended canonical transformation.
The underlying extended generating function $\FC_{2}^{\mu}$ was set up
to yield the required transformation rule for the connection coefficients.
The latter act as ``gauge coefficients'' that render globally (Lorentz-)invariant
systems form-invariant under the corresponding \emph{local\/} transformation group,
i.e.\ the diffeomorphism group.
The form-invariant extended Hamiltonian could then be deduced from the general
rules for extended canonical transformations in the realm of field theory.

As was noted by C.~Lanczos~\cite{lanczos38}, the Lagrangian of general relativity
should be a \emph{quadratic\/} function of the curvature tensor elements
in order to obey the \emph{principle of scale-independence}, which means that the
value of the action integral~(\ref{action-int3}) ``should not depend on the
arbitrary units employed in measuring the lengths of the space-time manifold.''
The Lagrangian~(\ref{eq:lag-genrel}) derived here meets this requirement.
Summarizing, the theory has the following properties
\begin{enumerate}
\item The space-time geometry is of Riemannian type, with a \emph{torsion\/}
of space-time not excluded \emph{a priori}.
Hence, the theory does not presuppose nor require the \emph{equivalence principle\/}---i.e.\
the equivalence of gravitational and inertial mass---to hold strictly.
\item The theory is based on the \emph{action principle}, which yields field equations
for the space-time and the connection coefficients and hence the metric tensor.
\item The \emph{general principle of relativity\/} is satisfied, hence, the theory is
\emph{form-invariant\/} under the \emph{canonical transformation\/} of the connection
coefficients and their conjugates---which ensures the action principle to be maintained.
This can be regarded as the realization of the gauge principle for a
diffeomorphism-invariant theory, with the connection coefficients playing
the role of ``gauge fields.''
\item The \emph{principle of scale invariance\/} holds, hence the theory is form-invariant
under transformations of the length scales of the space-time manifold.
\item The theory is \emph{unambiguous\/} in the sense that no other functions
of the Riemann curvature tensor emerge from the canonical transformation formalism.
\item In contrast to standard general relativity that is based on the
\emph{postulated\/} Einstein-Hilbert action, the Lagrangian~(\ref{eq:Lag-conn-coeff2})
is \emph{derived\/}, based on the requirement of its form-invariance under the canonical
transformation of the connection coefficients, and hence on the general principle of relativity.
Moreover, no \emph{ad hoc} assumption concerning the relation of the
connection coefficients with the metric is incorporated into the formalism.
Instead, it is the canonical equation (\ref{eq:feq-conn-coeff3}) that provides
the correlation of the connection coefficients with the metric.
\item For the source-free case ($\LC_{\mathrm{M}}\equiv0$), the theory is compatible with standard
general relativity in the torsion-free limit as it possesses then the Schwarzschild metric as a solution~\cite{stephenson58}.
For $\LC_{\mathrm{M}}\neq0$, the solution is expected to differ from that of standard
general relativity---especially if a torsion of space-time is not to be excluded \emph{a priori}.
\item A quantized theory that is based on the Lagrangian~(\ref{eq:lag-genrel}) is
\emph{renormalizable}~\cite{stelle77}.
It is a well-known fact that the energy of a gravitational field is \emph{not localizable}, hence
that the energy-momentum (pseudo-)tensor of the gravitational field depends on the frame of reference.
In a quantized theory of gravity, this would mean that the hypothetical interaction bosons (the ``gravitons'')
are the quanta of a nontensorial classical ``field'' that is represented by the connection coefficients
$\Gamma\indices{^{\xi}_{\mu\nu}}$.
\end{enumerate}
Remarkably, a Lagrangian of the form of Eq.~(\ref{eq:higher-curvature-lag}) that is quadratic in the
curvature tensor was already proposed by A.~Einstein in a personal letter to Hermann Weyl,
dated March 08, 1918~\cite{einstein18}, reasoning analogies with other classical field theories.
\appendix
\section{USEFUL IDENTITIES\label{app1}}
In order to show that the conventional Lagrangian $\LC$ description
of a dynamical system is compatible with the corresponding description
in terms of extended Lagrangians, we must make use of the following identities
\begin{eqnarray}
\pfrac{\phi_{I}}{x^{\alpha}}&=&\pfrac{\phi_{I}}{y^{i}}\pfrac{y^{i}}{x^{\alpha}}\nonumber\\
&&\Rightarrow\pfrac{\left(\pfrac{\phi_{I}}{x^{\alpha}}\right)}
{\left(\pfrac{\phi_{I}}{y^{\nu}}\right)}=\delta_{i}^{\nu}
\pfrac{y^{i}}{x^{\alpha}}=\pfrac{y^{\nu}}{x^{\alpha}}\nonumber\\
&&\Rightarrow\pfrac{\left(\pfrac{\phi_{I}}{x^{\alpha}}\right)}
{\left(\pfrac{y^{\nu}}{x^{\mu}}\right)}=
\pfrac{\phi_{I}}{y^{i}}\delta_{\nu}^{i}\delta_{\alpha}^{\mu}=
\pfrac{\phi_{I}}{y^{\nu}}\delta_{\alpha}^{\mu}\nonumber\\
\pfrac{x^{\nu}}{y^{k}}\pfrac{y^{k}}{x^{\mu}}&=&\delta_{\mu}^{\nu}\nonumber\\
&&\Rightarrow\pfrac{\left(\pfrac{y^{\nu}}{x^{\mu}}\right)}
{\left(\pfrac{x^{\alpha}}{y^{\beta}}\right)}=-\pfrac{y^{\nu}}{x^{\alpha}}
\pfrac{y^{\beta}}{x^{\mu}}\nonumber\\
&&\Rightarrow\ppfrac{y^{k}}{x^{\mu}}{x^{\xi}}\pfrac{x^{\nu}}{y^{k}}
=-\ppfrac{x^{\nu}}{y^{j}}{y^{k}}\pfrac{y^{j}}{x^{\xi}}\pfrac{y^{k}}{x^{\mu}}\nonumber\\
\pfrac{\left(\pfrac{\phi_{I}}{x^{\alpha}}\right)}
{\left(\pfrac{x^{\mu}}{y^{\nu}}\right)}&=&
\pfrac{\left(\pfrac{\phi_{I}}{x^{\alpha}}\right)}
{\left(\pfrac{y^{j}}{x^{i}}\right)}
\pfrac{\left(\pfrac{y^{j}}{x^{i}}\right)}
{\left(\pfrac{x^{\mu}}{y^{\nu}}\right)}\nonumber\\
&=&-\pfrac{\phi_{I}}{y^{j}}\pfrac{y^{j}}{x^{\mu}}\pfrac{y^{\nu}}{x^{\alpha}}
=-\pfrac{\phi_{I}}{x^{\mu}}\pfrac{y^{\nu}}{x^{\alpha}}.
\label{L1-identity}
\end{eqnarray}
Frequently, the derivative of the Jacobi determinant with
respect to the space-time coefficients needs to be inserted.
This quantity is easiest calculated on the basis of the
general formula for a determinant of an $n\times n$ matrix $A=(a_{ik})$.
With $S_{n}$ the set of all permutations $\pi$ of the
numbers $1,\ldots,n$, the determinant is given by the sum over all $\pi\in S_{n}$
\begin{displaymath}
\det A=\sum_{\pi\in S_{n}}\sgn\pi\;a_{1\pi(1)}\ldots a_{n\pi(n)}.
\end{displaymath}
Herein, $\sgn\pi=\pm1$ depending on the permutation to consist of
an \emph{even\/} or \emph{odd\/} number of elementary transpositions of pairs of numbers.
In the first case, $\sgn\pi=+1$ whereas $\sgn\pi=-1$ in the latter.
This means for $\det\Lambda$
\begin{displaymath}
\det\Lambda=\sum_{\pi\in S_{n+1}}\sgn\pi\;\pfrac{x^{0}}{y^{\pi(0)}}\ldots\pfrac{x^{n}}{y^{\pi(n)}}.
\end{displaymath}
With the sum over all $\alpha$, the expression
\begin{eqnarray*}
&&\pfrac{\det\Lambda}{\left(\pfrac{x^{\mu}}{y^{\alpha}}\right)}\pfrac{x^{\nu}}{y^{\alpha}}\\
&=&\Bigg(\sum_{\pi\in S_{n+1}}\sgn\pi\pfrac{x^{0}}{y^{\pi(0)}}
\ldots\pfrac{x^{\mu-1}}{y^{\pi(\mu-1)}}\delta_{\pi(\mu)}^{\alpha}\\
&&\qquad\qquad\qquad\times\pfrac{x^{\mu+1}}{y^{\pi(\mu+1)}}
\ldots\pfrac{x^{n}}{y^{\pi(n)}}\Bigg)\pfrac{x^{\nu}}{y^{\alpha}}\\
&=&\!\!\!\!\!\!\sum_{\pi\in S_{n+1}}\!\!\!\!\!\sgn\pi\pfrac{x^{0}}{y^{\pi(0)}}\!
\ldots\!\pfrac{x^{\mu-1}}{y^{\pi(\mu-1)}}
\pfrac{x^{\nu}}{y^{\pi(\mu)}}\pfrac{x^{\mu+1}}{y^{\pi(\mu+1)}}\!\ldots\!\pfrac{x^{n}}{y^{\pi(n)}}
\end{eqnarray*}
is either zero for $\nu\neq\mu$ as the derivative of $x^{\nu}$ then occurs twice or
equal to $\det\Lambda$ for $\nu=\mu$ as the derivative of $x^{\mu}$ is recovered
and thus yields the initial expression for the determinant $\det\Lambda$.
Thus
\begin{displaymath}
\pfrac{\det\Lambda}{\left(\pfrac{x^{\mu}}{y^{\alpha}}\right)}
\pfrac{x^{\nu}}{y^{\alpha}}=\delta_{\mu}^{\nu}\det\Lambda,
\end{displaymath}
hence
\begin{equation}\label{eq:detLambda-identity}
\pfrac{\det\Lambda}{\left(\pfrac{x^{\mu}}{y^{\nu}}\right)}=
\pfrac{y^{\nu}}{x^{\mu}}\det\Lambda.
\end{equation}
The correlation~(\ref{L1-def}) of the trivial extended Lagrangian $\LC_{\e}$
and conventional Lagrangian $\LC$ emerges from the requirement of
Eq.~(\ref{action-int1}) to yield the identical action $S$, hence to
describe the same physical system.
The derivative of a \emph{trivial\/} extended Lagrangian with respect to
the space-time coefficients yields the canonical energy-momentum tensor.
Explicitly,
\begin{eqnarray}
\pfrac{\LC_{\e}^{\mathrm{triv}}}{\left(\pfrac{x^{\mu}}{y^{\nu}}\right)}&=&
\LC\,\pfrac{\det\Lambda}{\left(\pfrac{x^{\mu}}{y^{\nu}}\right)}+
\pfrac{\LC}{\left(\pfrac{\phi_{I}}{x^{\alpha}}\right)}
\pfrac{\left(\pfrac{\phi_{I}}{x^{\alpha}}\right)}
{\left(\pfrac{x^{\mu}}{y^{\nu}}\right)}\det\Lambda\nonumber\\
&=&\LC\,\pfrac{y^{\nu}}{x^{\mu}}\det\Lambda
-\pfrac{\LC}{\left(\pfrac{\phi_{I}}{x^{\alpha}}\right)}
\pfrac{\phi_{I}}{x^{\mu}}\pfrac{y^{\nu}}{x^{\alpha}}\det\Lambda\nonumber\\
&=&\left(\delta_{\mu}^{i}\LC-\pfrac{\LC}{\left(\pfrac{\phi_{I}}{x^{\alpha}}\right)}
\pfrac{\phi_{I}}{x^{\mu}}\right)\pfrac{y^{\nu}}{x^{\alpha}}\det\Lambda\nonumber\\
&=&-\theta\indices{_{\mu}^{\alpha}}(x)\,\pfrac{y^{\nu}}{x^{\alpha}}\det\Lambda
=-\tilde{\theta}\indices{_{\mu}^{\alpha}}(x)\,\pfrac{y^{\nu}}{x^{\alpha}},
\label{L1-deri}
\end{eqnarray}
where $\theta\indices{_{\mu}^{\nu}}$ denotes the energy-momentum
tensor, and $\tilde{\theta}\indices{_{\mu}^{\nu}}$ the corresponding
\emph{tensor density\/} at the same space-time location.

A frequently used identity follows as
\begin{eqnarray}
&&\pfrac{}{y^{\alpha}}\pfrac{\det\Lambda}{\left(\pfrac{x^{\mu}}{y^{\alpha}}\right)}=
\pfrac{}{y^{\alpha}}\left(\pfrac{y^{\alpha}}{x^{\mu}}\det\Lambda\right)\nonumber\\
&=&\ppfrac{y^{\alpha}}{x^{\mu}}{x^{\beta}}\pfrac{x^{\beta}}{y^{\alpha}}\det\Lambda+
\pfrac{y^{\alpha}}{x^{\mu}}\pfrac{\det\Lambda}{\left(\pfrac{x^{i}}{y^{j}}\right)}
\ppfrac{x^{i}}{y^{j}}{y^{\alpha}}\nonumber\\
&=&\Bigg(\ppfrac{y^{\alpha}}{x^{\mu}}{x^{\beta}}\pfrac{x^{\beta}}{y^{\alpha}}+
\underbrace{\pfrac{y^{\alpha}}{x^{\mu}}\pfrac{y^{j}}{x^{i}}
\ppfrac{x^{i}}{y^{j}}{y^{\alpha}}}_{\stackrel{\mathrm{(\ref{L1-identity})}}{=}-\ppfrac{y^{\alpha}}{x^{\mu}}{x^{\beta}}
\pfrac{x^{\beta}}{y^{\alpha}}}\Bigg)\det\Lambda\nonumber\\
&\equiv&0.
\label{eq:ext-lag-identity}
\end{eqnarray}
This identity is used for setting up the extended set of
Euler-Lagrange equations.

The derivative of the contravariant metric with respect
to the space-time coefficients follows as
\begin{eqnarray}
&&\pfrac{g^{\mu\nu}(y)}{\left(\pfrac{x^{\alpha}}{y^{\beta}}\right)}=
g^{ij}(x)\left[\pfrac{\left(\pfrac{y^{\mu}}{x^{i}}\right)}{\left(\pfrac{x^{\alpha}}{y^{\beta}}\right)}
\pfrac{y^{\nu}}{x^{j}}+\pfrac{y^{\mu}}{x^{i}}\pfrac{\left(
\pfrac{y^{\nu}}{x^{j}}\right)}{\left(\pfrac{x^{\alpha}}{y^{\beta}}\right)}\right]\nonumber\\
&=&-g^{ab}(y)\pfrac{x^{i}}{y^{a}}\pfrac{x^{j}}{y^{b}}\left(
\pfrac{y^{\mu}}{x^{\alpha}}\pfrac{y^{\beta}}{x^{i}}\pfrac{y^{\nu}}{x^{j}}+
\pfrac{y^{\mu}}{x^{i}}\pfrac{y^{\nu}}{x^{\alpha}}\pfrac{y^{\beta}}{x^{j}}\right)\nonumber\\
&=&-g^{\beta\nu}(y)\pfrac{y^{\mu}}{x^{\alpha}}-g^{\mu\beta}(y)\pfrac{y^{\nu}}{x^{\alpha}}\nonumber\\
&=&-\left(\delta_{j}^{\mu}\,g^{\beta\nu}(y)+\delta_{j}^{\nu}\,g^{\mu\beta}(y)\right)\pfrac{y^{j}}{x^{\alpha}}.
\label{eq:deri-cov-metric3}
\end{eqnarray}
The derivative of the covariant metric with respect
to the space-time coefficients is then
\begin{eqnarray}
\pfrac{g_{\mu\nu}(y)}{\left(\pfrac{x^{\alpha}}{y^{\beta}}\right)}&=&
g_{ij}(x)\left[\pfrac{\left(\pfrac{x^{i}}{y^{\mu}}\right)}{\left(\pfrac{x^{\alpha}}{y^{\beta}}\right)}
\pfrac{x^{j}}{y^{\nu}}+\pfrac{x^{i}}{y^{\mu}}\pfrac{\left(
\pfrac{x^{j}}{y^{\nu}}\right)}{\left(\pfrac{x^{\alpha}}{y^{\beta}}\right)}\right]\nonumber\\
&=&g_{ab}(y)\pfrac{y^{a}}{x^{i}}\pfrac{y^{b}}{x^{j}}
\left(\delta_{\alpha}^{i}\delta_{\mu}^{\beta}\pfrac{x^{j}}{y^{\nu}}+
\pfrac{x^{i}}{y^{\mu}}\delta_{\alpha}^{j}\delta_{\nu}^{\beta}\right)\nonumber\\
&=&\left(\delta_{\mu}^{\beta}\,g_{j\nu}(y)+\delta_{\nu}^{\beta}\,g_{\mu j}(y)\right)
\pfrac{y^{j}}{x^{\alpha}}.
\label{eq:deri-cov-metric4}
\end{eqnarray}
\section{CONNECTION COEFFICIENTS\label{app2}}
The derivatives of a vector $a_{\mu}$ do \emph{not\/} transform as tensors
\begin{equation}\label{eq:nontensor-trans}
\pfrac{a_{\mu}(X)}{X^{\nu}}=\pfrac{a_{i}(x)}{x^{j}}\pfrac{x^{j}}{X^{\nu}}
\pfrac{x^{i}}{X^{\mu}}+a_{i}(x)\ppfrac{x^{i}}{X^{\mu}}{X^{\nu}},
\end{equation}
provided that the reference system $x(y)$ is curved with respect to the reference
system $X(y)$, which means that not all \emph{second derivatives\/} of the
$x^{\mu}$ in~(\ref{eq:nontensor-trans}) vanish.

Equation~(\ref{eq:nontensor-trans}) can be converted into a tensor equation
by introducing \emph{connection coefficients\/}
$\gamma\indices{^{\mu}_{\alpha\beta}}(x)$ and $\Gamma\indices{^{\mu}_{\alpha\beta}}(X)$
\begin{eqnarray*}
\pfrac{a_{\mu}(X)}{X^{\nu}}&=&\pfrac{x^{j}}{X^{\nu}}\pfrac{x^{i}}{X^{\mu}}\pfrac{a_{i}(x)}{x^{j}}+
a_{k}(x)\ppfrac{x^{k}}{X^{\mu}}{X^{\nu}}\\
&=&\pfrac{x^{j}}{X^{\nu}}\pfrac{x^{i}}{X^{\mu}}\left(
\pfrac{a_{i}(x)}{x^{j}}-a_{k}(x)\gamma\indices{^{k}_{ij}}(x)\right)\\
&&\mbox{}+a_{k}(X)\,\Gamma\indices{^{k}_{\mu\nu}}(X).
\end{eqnarray*}
Then
\begin{eqnarray*}
&&\pfrac{a_{\mu}(X)}{X^{\nu}}-a_{k}(X)\Gamma\indices{^{k}_{\mu\nu}}(X)\\
&&\quad=\pfrac{x^{j}}{X^{\nu}}\pfrac{x^{i}}{X^{\mu}}\left(
\pfrac{a_{i}(x)}{x^{j}}-a_{k}(x)\gamma\indices{^{k}_{ij}}(x)\right),
\end{eqnarray*}
which shows that the quantity
\begin{equation}\label{eq:def-cov-deri1}
a_{i;j}\equiv\pfrac{a_{i}(x)}{x^{j}}-a_{k}(x)\gamma\indices{^{k}_{ij}}(x)
\end{equation}
transforms as a tensor, provided that the connection coefficients transform as
\begin{eqnarray*}
a_{k}(x)\ppfrac{x^{k}}{X^{\mu}}{X^{\nu}}&=&
a_{k}(x)\pfrac{x^{k}}{X^{j}}\Gamma\indices{^{j}_{\mu\nu}}(X)\\
&&\mbox{}-\pfrac{x^{j}}{X^{\nu}}\pfrac{x^{i}}{X^{\mu}}a_{k}(x)\gamma\indices{^{k}_{ij}}(x).
\end{eqnarray*}
As this equation holds for arbitrary $a_{k}(x)$, it follows that
\begin{equation}\label{eq:christoffel-trans1}
\Gamma\indices{^{j}_{\mu\nu}}(X)\pfrac{x^{k}}{X^{j}}=
\gamma\indices{^{k}_{ij}}(x)\pfrac{x^{i}}{X^{\mu}}
\pfrac{x^{j}}{X^{\nu}}+\ppfrac{x^{k}}{X^{\mu}}{X^{\nu}},
\end{equation}
and finally after contraction with $\partial X^{\alpha}/\partial x^{k}$,
\begin{equation}\label{eq:christoffel-trans}
\Gamma\indices{^{\alpha}_{\mu\nu}}(X)=
\gamma\indices{^{k}_{ij}}(x)\pfrac{x^{i}}{X^{\mu}}\pfrac{x^{j}}{X^{\nu}}
\pfrac{X^{\alpha}}{x^{k}}+\ppfrac{x^{k}}{X^{\mu}}{X^{\nu}}\pfrac{X^{\alpha}}{x^{k}}.
\end{equation}
This equation provides the \emph{unique\/} correlation of the space-time
coefficients and their derivatives with the connection coefficients.
The connection coefficients are symmetric in their lower indices for torsion-free space.
Otherwise, their skew-symmetric part define the \emph{torsion tensor}.
\section{EULER-LAGRANGE EQUATIONS FOR SPACE-TIME COEFFICIENTS AND CONNECTION COEFFICIENTS\label{app3}}
Given an extended Lagrangian that depends on the space-time event,
the connection coefficients, and their respective space-time derivatives.
The action functional over a space-time region $R$ is then
\begin{equation}\label{action-int3a}
S=\int_{R}\LC_{\e}\!\left(
\gamma\indices{^{\eta}_{\alpha\xi}}(y),\pfrac{\gamma\indices{^{\eta}_{\alpha\xi}}}{y^{\nu}},
x^{\mu}(y),\pfrac{x^{\mu}}{y^{\nu}}\right)\!\dd^{4}y,\quad\delta S\stackrel{!}{=}0.
\end{equation}
The variation of $\LC_{\e}$ follows as
\begin{eqnarray*}
\delta\LC_{\e}=\pfrac{\LC_{\e}}{x^{\alpha}}\delta x^{\alpha}&+&
\pfrac{\LC_{\e}}{\left(\pfrac{x^{\alpha}}{y^{\beta}}\right)}\delta\left(\pfrac{x^{\alpha}}{y^{\beta}}\right)+
\pfrac{\LC_{\e}}{\gamma\indices{^{\eta}_{\alpha\xi}}}\delta\gamma\indices{^{\eta}_{\alpha\xi}}\\
&+&\pfrac{\LC_{\e}}{\left(\pfrac{\gamma\indices{^{\eta}_{\alpha\xi}}}{y^{\beta}}\right)}
\delta\left(\pfrac{\gamma\indices{^{\eta}_{\alpha\xi}}}{y^{\beta}}\right).
\end{eqnarray*}
As the independent variables $y^{\mu}$ are not varied, the differentiation with respect
to $y^{\beta}$ may be interchanged with the variation.
This yields the equivalent representations of $\delta\LC_{\e}$
\begin{eqnarray*}
\delta\LC_{\e}=\pfrac{\LC_{\e}}{x^{\alpha}}\delta x^{\alpha}&+&
\pfrac{\LC_{\e}}{\left(\pfrac{x^{\alpha}}{y^{\beta}}\right)}\pfrac{\left(\delta x^{\alpha}\right)}{y^{\beta}}+
\pfrac{\LC_{\e}}{\gamma\indices{^{\eta}_{\alpha\xi}}}\delta\gamma\indices{^{\eta}_{\alpha\xi}}\\
&+&\pfrac{\LC_{\e}}{\left(\pfrac{\gamma\indices{^{\eta}_{\alpha\xi}}}{y^{\beta}}\right)}
\pfrac{\left(\delta\gamma\indices{^{\eta}_{\alpha\xi}}\right)}{y^{\beta}}
\end{eqnarray*}
and
\begin{eqnarray*}
\delta\LC_{\e}&=&\!\left(\!\pfrac{\LC_{\e}}{x^{\alpha}}\!-\!\pfrac{}{y^{\beta}}
\pfrac{\LC_{\e}}{\left(\pfrac{x^{\alpha}}{y^{\beta}}\right)}\!\right)\!\delta x^{\alpha}\!+\!
\pfrac{}{y^{\beta}}\!\left(\!\pfrac{\LC_{\e}}{\left(\pfrac{x^{\alpha}}{y^{\beta}}\right)}\delta x^{\alpha}\!\right)\\
&&\mbox{}+\left(\pfrac{\LC_{\e}}{\gamma\indices{^{\eta}_{\alpha\xi}}}-\pfrac{}{y^{\beta}}
\pfrac{\LC_{\e}}{\left(\pfrac{\gamma\indices{^{\eta}_{\alpha\xi}}}{y^{\beta}}\right)}\right)
\delta\gamma\indices{^{\eta}_{\alpha\xi}}\\
&&\mbox{}+\pfrac{}{y^{\beta}}\left(
\pfrac{\LC_{\e}}{\left(\pfrac{\gamma\indices{^{\eta}_{\alpha\xi}}}{y^{\beta}}\right)}
\delta\gamma\indices{^{\eta}_{\alpha\xi}}\right).
\end{eqnarray*}
According to Gauss' theorem, the divergence terms can be converted into surface
terms in the action functional.
This means explicitly
\begin{eqnarray*}
\int_{R}\pfrac{}{y^{\beta}}\left(\pfrac{\LC_{\e}}{\left(
\pfrac{x^{\alpha}}{y^{\beta}}\right)}\delta x^{\alpha}\right)\dd^{4}y&=&
\oint_{\partial R}\pfrac{\LC_{\e}}{\left(
\pfrac{x^{\alpha}}{y^{\beta}}\right)}\delta x^{\alpha}\dd S_{\beta}\\
\int_{R}\!\pfrac{}{y^{\beta}}\!\!\left(\!\!
\pfrac{\LC_{\e}}{\left(\pfrac{\gamma\indices{^{\eta}_{\alpha\xi}}}{y^{\beta}}\right)}
\delta\gamma\indices{^{\eta}_{\alpha\xi}}\!\!\right)\!\!\!\dd^{4}y\!&=&\!
\oint_{\partial R}\!\pfrac{\LC_{\e}}{\left(\pfrac{\gamma\indices{^{\eta}_{\alpha\xi}}}{y^{\beta}}\right)}
\delta\gamma\indices{^{\eta}_{\alpha\xi}}\!\dd S_{\beta}
\end{eqnarray*}
with $\dd S_{\beta}$ denoting the $\beta$ component of the normal vector
on the boundary surface $\partial R$ of the volume $R$.
Both surface integrals vanish since on the boundary of the space-time region $R$
\begin{displaymath}
\left.\delta x^{\alpha}\right|_{\partial R}=0,\qquad
\left.\delta\gamma\indices{^{\eta}_{\alpha\xi}}\right|_{\partial R}=0,
\end{displaymath}
which implies that the space-time geometry is flat on $\partial R$.
The action principle $\delta S\stackrel{!}{=}0$ then reduces to
\begin{eqnarray*}
0\stackrel{!}{=}\int_{R}&&\Bigg[\Bigg(\pfrac{\LC_{\e}}{x^{\alpha}}-\pfrac{}{y^{\beta}}
\pfrac{\LC_{\e}}{\left(\pfrac{x^{\alpha}}{y^{\beta}}\right)}\Bigg)\delta x^{\alpha}\\
&&\qquad\mbox{}+
\Bigg(\pfrac{\LC_{\e}}{\gamma\indices{^{\eta}_{\alpha\xi}}}-\pfrac{}{y^{\beta}}
\pfrac{\LC_{\e}}{\left(\pfrac{\gamma\indices{^{\eta}_{\alpha\xi}}}{y^{\beta}}\right)}\Bigg)
\delta\gamma\indices{^{\eta}_{\alpha\xi}}\Bigg]\dd^{4}y
\end{eqnarray*}
As the variations $\delta x^{\alpha}$ and $\delta\gamma\indices{^{\eta}_{\alpha\xi}}$
are arbitrary and mutually independent by assumption, this condition can
only be satisfied if the expressions in parentheses vanish simultaneously
\begin{displaymath}
\pfrac{}{y^{\beta}}\pfrac{\LC_{\e}}{\left(\pfrac{x^{\alpha}}{y^{\beta}}\right)}-
\pfrac{\LC_{\e}}{x^{\alpha}}=0,\quad\pfrac{}{y^{\beta}}
\pfrac{\LC_{\e}}{\left(\pfrac{\gamma\indices{^{\eta}_{\alpha\xi}}}{y^{\beta}}\right)}-
\pfrac{\LC_{\e}}{\gamma\indices{^{\eta}_{\alpha\xi}}}=0.
\end{displaymath}
These equations are the Euler-Lagrange equations for the space-time
and the connection coefficients.
\section{EXPLICIT CALCULATION OF THE TRANSFORMATION RULE (\ref{eq:ham-conn-coeff1})\label{app4}}
In expanded form, Eq.~(\ref{eq:ham-conn-coeff}) reads
\begin{widetext}
\begin{eqnarray}
\tilde{T}\indices{_{\alpha}^{\beta}}\pfrac{X^{\alpha}}{y^{\beta}}-
\tilde{t}\indices{_{\alpha}^{\beta}}\pfrac{x^{\alpha}}{y^{\beta}}=
\tilde{R}\indices{_{\eta}^{\alpha\xi\mu}}&&\left[\gamma\indices{^{k}_{ij}}\left(
\ppfrac{X^{\eta}}{x^{k}}{x^{\nu}}\pfrac{x^{\nu}}{X^{\mu}}\pfrac{x^{i}}{X^{\alpha}}\pfrac{x^{j}}{X^{\xi}}+
\ppfrac{x^{i}}{X^{\alpha}}{X^{\mu}}\pfrac{X^{\eta}}{x^{k}}\pfrac{x^{j}}{X^{\xi}}+
\ppfrac{x^{j}}{X^{\xi}}{X^{\mu}}\pfrac{X^{\eta}}{x^{k}}\pfrac{x^{i}}{X^{\alpha}}\right)\right.\nonumber\\
&&\qquad\quad\left.\mbox{}+\ppfrac{X^{\eta}}{x^{k}}{x^{\nu}}\ppfrac{x^{k}}{X^{\alpha}}{X^{\xi}}
\pfrac{x^{\nu}}{X^{\mu}}+\pppfrac{x^{k}}{X^{\alpha}}{X^{\xi}}{X^{\mu}}\pfrac{X^{\eta}}{x^{k}}\right].
\label{eq:ham-conn-coeff2}
\end{eqnarray}
This expression is now split into a skew-symmetric and a symmetric part
of $\tilde{R}\indices{_{\eta}^{\alpha\xi\mu}}$ in the indices $\xi,\mu$ according to
\begin{displaymath}
\tilde{R}\indices{_{\eta}^{\alpha\xi\mu}}=
\onehalf\left(\tilde{R}\indices{_{\eta}^{\alpha\xi\mu}}-\tilde{R}\indices{_{\eta}^{\alpha\mu\xi}}\right)+
\onehalf\left(\tilde{R}\indices{_{\eta}^{\alpha\xi\mu}}+\tilde{R}\indices{_{\eta}^{\alpha\mu\xi}}\right)=
\tilde{R}\indices{_{\eta}^{\alpha[\xi\mu]}}+\tilde{R}\indices{_{\eta}^{\alpha(\xi\mu)}}.
\end{displaymath}
For the skew-symmetric part, $\tilde{R}\indices{_{\eta}^{\alpha[\xi\mu]}}$,
the two terms in~(\ref{eq:ham-conn-coeff2}) symmetric in $\xi,\mu$ vanish, hence
\begin{eqnarray*}
&&\tilde{R}\indices{_{\eta}^{\alpha[\xi\mu]}}\left[\gamma\indices{^{k}_{ij}}\left(
\ppfrac{X^{\eta}}{x^{k}}{x^{\nu}}\pfrac{x^{\nu}}{X^{\mu}}\pfrac{x^{i}}{X^{\alpha}}\pfrac{x^{j}}{X^{\xi}}+
\ppfrac{x^{i}}{X^{\alpha}}{X^{\mu}}\pfrac{X^{\eta}}{x^{k}}\pfrac{x^{j}}{X^{\xi}}\right)+
\ppfrac{X^{\eta}}{x^{k}}{x^{\nu}}\pfrac{x^{\nu}}{X^{\mu}}\ppfrac{x^{k}}{X^{\alpha}}{X^{\xi}}\right]\\
&=&\tilde{R}\indices{_{\eta}^{\alpha[\xi\mu]}}\left[
\ppfrac{X^{\eta}}{x^{k}}{x^{\nu}}\pfrac{x^{\nu}}{X^{\mu}}\left(
\gamma\indices{^{k}_{ij}}\pfrac{x^{i}}{X^{\alpha}}\pfrac{x^{j}}{X^{\xi}}+\ppfrac{x^{k}}{X^{\alpha}}{X^{\xi}}\right)+
\gamma\indices{^{k}_{ij}}\ppfrac{x^{i}}{X^{\alpha}}{X^{\mu}}\pfrac{X^{\eta}}{x^{k}}\pfrac{x^{j}}{X^{\xi}}\right]\\
&=&\tilde{R}\indices{_{\eta}^{\alpha[\xi\mu]}}\left[\Gamma\indices{^{j}_{\alpha\xi}}
\ppfrac{X^{\eta}}{x^{k}}{x^{\nu}}\pfrac{x^{k}}{X^{j}}\pfrac{x^{\nu}}{X^{\mu}}+
\gamma\indices{^{k}_{ij}}\ppfrac{x^{i}}{X^{\alpha}}{X^{\mu}}\pfrac{X^{\eta}}{x^{k}}\pfrac{x^{j}}{X^{\xi}}\right]\\
&=&\tilde{R}\indices{_{\eta}^{\alpha[\xi\mu]}}\left[\Gamma\indices{^{j}_{\alpha\xi}}\left(
\gamma\indices{^{i}_{k\nu}}\pfrac{X^{\eta}}{x^{i}}\pfrac{x^{k}}{X^{j}}\pfrac{x^{\nu}}{X^{\mu}}-
\Gamma\indices{^{\eta}_{j\mu}}\right)+\gamma\indices{^{k}_{ij}}\left(\Gamma\indices{^{a}_{\alpha\mu}}
\pfrac{x^{i}}{X^{a}}-\gamma\indices{^{i}_{ab}}\pfrac{x^{a}}{X^{\alpha}}\pfrac{x^{b}}{X^{\mu}}
\right)\pfrac{X^{\eta}}{x^{k}}\pfrac{x^{j}}{X^{\xi}}\right]\\
&=&\tilde{R}\indices{_{\eta}^{\alpha[\xi\mu]}}\left(
-\Gamma\indices{^{i}_{\alpha\xi}}\Gamma\indices{^{\eta}_{i\mu}}-
\gamma\indices{^{i}_{ab}}\gamma\indices{^{k}_{ij}}\pfrac{x^{a}}{X^{\alpha}}\pfrac{x^{b}}{X^{\mu}}
\pfrac{X^{\eta}}{x^{k}}\pfrac{x^{j}}{X^{\xi}}+\Gamma\indices{^{j}_{\alpha\xi}}\gamma\indices{^{i}_{k\nu}}
\pfrac{X^{\eta}}{x^{i}}\pfrac{x^{k}}{X^{j}}\pfrac{x^{\nu}}{X^{\mu}}+
\Gamma\indices{^{j}_{\alpha\mu}}\gamma\indices{^{i}_{k\nu}}
\pfrac{X^{\eta}}{x^{i}}\pfrac{x^{k}}{X^{j}}\pfrac{x^{\nu}}{X^{\xi}}\right)\\
&=&-\tilde{R}\indices{_{\eta}^{\alpha[\xi\mu]}}\Gamma\indices{^{i}_{\alpha\xi}}\Gamma\indices{^{\eta}_{i\mu}}+
\gamma\indices{^{i}_{ab}}\gamma\indices{^{k}_{ij}}\tilde{R}\indices{_{\eta}^{\alpha[\xi\mu]}}
\pfrac{x^{a}}{X^{\alpha}}\pfrac{x^{b}}{X^{\xi}}\pfrac{X^{\eta}}{x^{k}}\pfrac{x^{j}}{X^{\mu}}\\
&=&-\tilde{R}\indices{_{\eta}^{\alpha[\xi\mu]}}\Gamma\indices{^{i}_{\alpha\xi}}\Gamma\indices{^{\eta}_{i\mu}}+
\tilde{r}\indices{_{k}^{a[bj]}}\gamma\indices{^{i}_{ab}}\gamma\indices{^{k}_{ij}}=
-\tilde{R}\indices{_{\eta}^{\alpha[\xi\mu]}}\Gamma\indices{^{i}_{\alpha\xi}}\Gamma\indices{^{\eta}_{i\mu}}+
\tilde{r}\indices{_{\eta}^{\alpha[\xi\mu]}}\gamma\indices{^{i}_{\alpha\xi}}\gamma\indices{^{\eta}_{i\mu}}\\
&=&-\onehalf\tilde{R}\indices{_{\eta}^{\alpha\xi\mu}}\left(
\Gamma\indices{^{i}_{\alpha\xi}}\Gamma\indices{^{\eta}_{i\mu}}-
\Gamma\indices{^{i}_{\alpha\mu}}\Gamma\indices{^{\eta}_{i\xi}}\right)+
\onehalf\tilde{r}\indices{_{\eta}^{\alpha\xi\mu}}\left(
\gamma\indices{^{i}_{\alpha\xi}}\gamma\indices{^{\eta}_{i\mu}}-
\gamma\indices{^{i}_{\alpha\mu}}\gamma\indices{^{\eta}_{i\xi}}\right).
\end{eqnarray*}
The two mixed terms in $\Gamma,\gamma$ cancel each other due to the skew-symmetry of
$\tilde{R}\indices{_{\eta}^{\alpha[\xi\mu]}}$ in $\xi,\mu$.
\end{widetext}
The contribution of~(\ref{eq:ham-conn-coeff}) emerging from the \emph{symmetric\/}
part $\tilde{R}\indices{_{\eta}^{\alpha(\xi\mu)}}$ can be expressed in terms
of the derivatives of the connection coefficients, whose transformation rule is
\begin{eqnarray*}
\pfrac{\Gamma\indices{^{\eta}_{\alpha\xi}}}{X^{\kappa}}\pfrac{X^{\kappa}}{x^{\nu}}&=&
\pfrac{\gamma\indices{^{k}_{ij}}}{x^{\nu}}\pfrac{X^{\eta}}{x^{k}}
\pfrac{x^{i}}{X^{\alpha}}\pfrac{x^{j}}{X^{\xi}}\\
&&\mbox{}+\gamma\indices{^{k}_{ij}}\pfrac{}{x^{\nu}}\left(\pfrac{X^{\eta}}{x^{k}}
\pfrac{x^{i}}{X^{\alpha}}\pfrac{x^{j}}{X^{\xi}}\right)\\
&&\mbox{}+\pfrac{}{x^{\nu}}\left(\pfrac{X^{\eta}}{x^{k}}\ppfrac{x^{k}}{X^{\alpha}}{X^{\xi}}\right).
\end{eqnarray*}
Thus
\begin{eqnarray*}
&&\tilde{R}\indices{_{\eta}^{\alpha(\xi\mu)}}\pfrac{x^{\nu}}{X^{\mu}}
\left[\gamma\indices{^{k}_{ij}}\pfrac{}{x^{\nu}}\left(
\pfrac{X^{\eta}}{x^{k}}\pfrac{x^{i}}{X^{\alpha}}\pfrac{x^{j}}{X^{\xi}}\right)\right.\\
&&\left.\qquad\qquad\qquad\mbox{}+
\pfrac{}{x^{\nu}}\left(\pfrac{X^{\eta}}{x^{k}}\ppfrac{x^{k}}{X^{\alpha}}{X^{\xi}}\right)\right]\\
&=&\tilde{R}\indices{_{\eta}^{\alpha(\xi\mu)}}\pfrac{x^{\nu}}{X^{\mu}}\left(
\pfrac{\Gamma\indices{^{\eta}_{\alpha\xi}}}{X^{\kappa}}\pfrac{X^{\kappa}}{x^{\nu}}-
\pfrac{\gamma\indices{^{k}_{ij}}}{x^{\nu}}\pfrac{X^{\eta}}{x^{k}}
\pfrac{x^{i}}{X^{\alpha}}\pfrac{x^{j}}{X^{\xi}}\right)\\
&=&\tilde{R}\indices{_{\eta}^{\alpha(\xi\mu)}}
\pfrac{\Gamma\indices{^{\eta}_{\alpha\xi}}}{X^{\mu}}-
\tilde{r}\indices{_{k}^{i(j\nu)}}\pfrac{\gamma\indices{^{k}_{ij}}}{x^{\nu}}\\
&=&\tilde{R}\indices{_{\eta}^{\alpha(\xi\mu)}}
\pfrac{\Gamma\indices{^{\eta}_{\alpha\xi}}}{X^{\mu}}-
\tilde{r}\indices{_{\eta}^{\alpha(\xi\mu)}}\pfrac{\gamma\indices{^{\eta}_{\alpha\xi}}}{x^{\mu}}\\
&=&\onehalf\tilde{R}\indices{_{\eta}^{\alpha\xi\mu}}\!\!\left(\!
\pfrac{\Gamma\indices{^{\eta}_{\alpha\xi}}}{X^{\mu}}\!+\!\pfrac{\Gamma\indices{^{\eta}_{\alpha\mu}}}{X^{\xi}}\!\right)\!-\!
\onehalf\tilde{r}\indices{_{\eta}^{\alpha\xi\mu}}\!\!\left(\!
\pfrac{\gamma\indices{^{\eta}_{\alpha\xi}}}{x^{\mu}}+\pfrac{\gamma\indices{^{\eta}_{\alpha\mu}}}{x^{\xi}}\!\right)\!.
\end{eqnarray*}
The total transformation rule~(\ref{eq:ham-conn-coeff2}) expressed in terms
of connection coefficients is then
\begin{eqnarray*}
&&\qquad\tilde{T}\indices{_{\alpha}^{\beta}}\pfrac{X^{\alpha}}{y^{\beta}}-
\tilde{t}\indices{_{\alpha}^{\beta}}\pfrac{x^{\alpha}}{y^{\beta}}\\
&&=\onehalf\tilde{R}\indices{_{\eta}^{\alpha\xi\mu}}\!\!\left(\!
\pfrac{\Gamma\indices{^{\eta}_{\alpha\xi}}}{X^{\mu}}+\pfrac{\Gamma\indices{^{\eta}_{\alpha\mu}}}{X^{\xi}}-
\Gamma\indices{^{i}_{\alpha\xi}}\Gamma\indices{^{\eta}_{i\mu}}+
\Gamma\indices{^{i}_{\alpha\mu}}\Gamma\indices{^{\eta}_{i\xi}}\!\right)\\
&&\quad\mbox{}-\onehalf\tilde{r}\indices{_{\eta}^{\alpha\xi\mu}}\left(
\pfrac{\gamma\indices{^{\eta}_{\alpha\xi}}}{x^{\mu}}+\pfrac{\gamma\indices{^{\eta}_{\alpha\mu}}}{x^{\xi}}-
\gamma\indices{^{i}_{\alpha\xi}}\gamma\indices{^{\eta}_{i\mu}}+
\gamma\indices{^{i}_{\alpha\mu}}\gamma\indices{^{\eta}_{i\xi}}\right).
\end{eqnarray*}
\section{EXPLICIT CALCULATION OF THE EULER-LAGRANGE EQUATION (\ref{eq:sec-order-einstein0})\label{app5}}
The form-invariant extended Lagrangian describing the source-free space-time dynamics is given by
\begin{eqnarray*}
&&\LC_{\e}\left(\gamma\indices{^{\eta}_{\alpha\xi}},\pfrac{\gamma\indices{^{\eta}_{\alpha\xi}}}{x^{\nu}},
\pfrac{x^{\mu}}{y^{\nu}}\right)\\
&&\qquad=-\quarter g_{\kappa\eta}g^{\beta\alpha}g^{\lambda\xi}g^{\zeta\tau}
r\indices{^{\kappa}_{\beta\lambda\zeta}}r\indices{^{\eta}_{\alpha\xi\tau}}\det\Lambda,
\end{eqnarray*}
with $g_{\kappa\eta}$ denoting the covariant metric tensor and
$g^{\beta\alpha}$ its contravariant counterpart.
In this description, the curvature tensor $r\indices{^{\kappa}_{\beta\lambda\zeta}}$---as
defined by Eq.~(\ref{eq:Riemann-tensor2})---only depends on the connection
coefficients and their space-time derivatives.
Thus, only the metric tensors and $\det\Lambda$ have derivatives with respect
to the space-time coefficients $\partial x^{\mu}/\partial y^{\nu}$.
These derivatives were worked out explicitly with Eqs.~(\ref{eq:deri-cov-metric3}),
(\ref{eq:deri-cov-metric4}), and (\ref{eq:detLambda-identity}).
The derivative of $\LC_{\e}$ with respect to the space-time coefficients follows as
\begin{widetext}
\begin{eqnarray*}
\pfrac{\LC_{\e}}{\left(\pfrac{x^{\mu}}{y^{\nu}}\right)}&=&-\quarter
r\indices{^{\kappa}_{\beta\lambda\zeta}}r\indices{^{\eta}_{\alpha\xi\tau}}\det\Lambda\left[
\pfrac{g_{\kappa\eta}(y)}{\left(\pfrac{x^{\mu}}{y^{\nu}}\right)}g^{\beta\alpha}g^{\lambda\xi}g^{\zeta\tau}+
\pfrac{g^{\beta\alpha}(y)}{\left(\pfrac{x^{\mu}}{y^{\nu}}\right)}g_{\kappa\eta}g^{\lambda\xi}g^{\zeta\tau}\right.\\
&&\left.\mbox{}+\pfrac{g^{\lambda\xi}(y)}{\left(\pfrac{x^{\mu}}{y^{\nu}}\right)}g_{\kappa\eta}g^{\beta\alpha}g^{\zeta\tau}+
\pfrac{g^{\zeta\tau}(y)}{\left(\pfrac{x^{\mu}}{y^{\nu}}\right)}g_{\kappa\eta}g^{\beta\alpha}g^{\lambda\xi}+
\pfrac{y^{\nu}}{x^{\mu}}g_{\kappa\eta}g^{\beta\alpha}g^{\lambda\xi}g^{\zeta\tau}\right]\\
&=&-\quarter r\indices{^{\kappa}_{\beta\lambda\zeta}}r\indices{^{\eta}_{\alpha\xi\tau}}\det\Lambda
\pfrac{y^{j}}{x^{\mu}}\left[\left(\delta_{\kappa}^{\nu}\,g_{j\eta}+\delta_{\eta}^{\nu}\,g_{\kappa j}
\vphantom{\delta_{j}^{\beta}}\right)g^{\beta\alpha}g^{\lambda\xi}g^{\zeta\tau}\right.\\
&&\mbox{}-
\left(\delta_{j}^{\beta}\,g^{\nu\alpha}+\delta_{j}^{\alpha}\,g^{\beta\nu}\right)g_{\kappa\eta}g^{\lambda\xi}g^{\zeta\tau}-
\left(\delta_{j}^{\lambda}\,g^{\nu\xi}+\delta_{j}^{\xi}\,g^{\lambda\nu}\right)g_{\kappa\eta}g^{\beta\alpha}g^{\zeta\tau}\\
&&\left.\mbox{}-
\left(\delta_{j}^{\zeta}\,g^{\nu\tau}+\delta_{j}^{\tau}\,g^{\zeta\nu}\right)g_{\kappa\eta}g^{\beta\alpha}g^{\lambda\xi}+
\delta_{j}^{\nu}g_{\kappa\eta}g^{\beta\alpha}g^{\lambda\xi}g^{\zeta\tau}\right]\\
&=&\quarter\det\Lambda\pfrac{y^{j}}{x^{\mu}}\left(\mbox{}-
r\indices{^{\nu}_{\beta\lambda\zeta}}r\indices{^{\eta}_{\alpha\xi\tau}}g_{j\eta}g^{\beta\alpha}g^{\lambda\xi}g^{\zeta\tau}-
r\indices{^{\kappa}_{\beta\lambda\zeta}}r\indices{^{\nu}_{\alpha\xi\tau}}g_{\kappa j}g^{\beta\alpha}g^{\lambda\xi}g^{\zeta\tau}\right.\\
&&\qquad\qquad\qquad\mbox{}+
r\indices{^{\kappa}_{j\lambda\zeta}}r\indices{^{\eta}_{\alpha\xi\tau}}g_{\kappa\eta}g^{\nu\alpha}g^{\lambda\xi}g^{\zeta\tau}+
r\indices{^{\kappa}_{\beta\lambda\zeta}}r\indices{^{\eta}_{j\xi\tau}}g_{\kappa\eta}g^{\beta\nu}g^{\lambda\xi}g^{\zeta\tau}\\
&&\qquad\qquad\qquad\mbox{}+
r\indices{^{\kappa}_{\beta j\zeta}}r\indices{^{\eta}_{\alpha\xi\tau}}g_{\kappa\eta}g^{\beta\alpha}g^{\nu\xi}g^{\zeta\tau}+
r\indices{^{\kappa}_{\beta\lambda\zeta}}r\indices{^{\eta}_{\alpha j\tau}}g_{\kappa\eta}g^{\beta\alpha}g^{\lambda\nu}g^{\zeta\tau}\\
&&\qquad\qquad\qquad\mbox{}+
r\indices{^{\kappa}_{\beta\lambda j}}r\indices{^{\eta}_{\alpha\xi\tau}}g_{\kappa\eta}g^{\beta\alpha}g^{\lambda\xi}g^{\nu\tau}+
r\indices{^{\kappa}_{\beta\lambda\zeta}}r\indices{^{\eta}_{\alpha\xi j}}g_{\kappa\eta}g^{\beta\alpha}g^{\lambda\xi}g^{\zeta\nu}\\
&&\qquad\qquad\qquad\left.\mbox{}-\delta_{j}^{\nu}
r\indices{^{\kappa}_{\beta\lambda\zeta}}r\indices{^{\eta}_{\alpha\xi\tau}}g_{\kappa\eta}g^{\beta\alpha}g^{\lambda\xi}g^{\zeta\tau}\right)\\
&=&\quarter\det\Lambda\pfrac{y^{j}}{x^{\mu}}\left(-2r^{\nu\alpha\xi\tau}r_{j\alpha\xi\tau}
+2r^{\eta\nu\xi\tau}r_{\eta j\xi\tau}+2r^{\eta\alpha\nu\tau}r_{\eta\alpha j\tau}+
2r^{\eta\alpha\xi\nu}r_{\eta\alpha\xi j}-\delta_{j}^{\nu}r^{\eta\alpha\xi\tau}r_{\eta\alpha\xi\tau}\right)\\
&=&\left.\left(-\onehalf r^{\nu\alpha\xi\tau}r_{j\alpha\xi\tau}+
\onehalf r^{\eta\nu\xi\tau}r_{\eta j\xi\tau}+r^{\eta\alpha\xi\nu}r_{\eta\alpha\xi j}-\quarter\delta_{j}^{\nu}
r^{\eta\alpha\xi\tau}r_{\eta\alpha\xi\tau}\right)\right|_{y}\pfrac{y^{j}}{x^{\mu}}\det\Lambda\\
&=&\left.\left(-\onehalf r^{j\alpha\xi\tau}r_{\mu\alpha\xi\tau}+
\onehalf r^{\eta j\xi\tau}r_{\eta\mu\xi\tau}+r^{\eta\alpha\xi j}r_{\eta\alpha\xi\mu}-\quarter\delta_{\mu}^{j}
r^{\eta\alpha\xi\tau}r_{\eta\alpha\xi\tau}\right)\right|_{x}\pfrac{y^{\nu}}{x^{j}}\det\Lambda.
\end{eqnarray*}
\end{widetext}
The first two terms cancel under the precondition that
$r^{\nu\alpha\xi\tau}$ is skew-symmetric also in its \emph{first\/} index pair.
This is ensured if a metric is present~(\cite[p~324]{misner}), which is assumed in the actual context.
The final result is thus
\begin{displaymath}
\pfrac{\LC_{\e}}{\left(\pfrac{x^{\mu}}{y^{\nu}}\right)}=\left(r^{\eta\alpha\xi j}\,r_{\eta\alpha\xi\mu}-
\quarter\delta_{\mu}^{j}r^{\eta\alpha\xi\beta}r_{\eta\alpha\xi\beta }\right)\pfrac{y^{\nu}}{x^{j}}\det\Lambda.
\end{displaymath}
%

\end{document}